\documentclass[final]{siamltex}

\usepackage{amsmath,amssymb,amsfonts,epsfig,pstricks,subfig,url,overpic,pst-plot,pst-node} 
\usepackage{rotating}
\newsavebox{\IBox}

\newcommand{\newrho}{\overline{\varrho}}
\newcommand{\newsigma}{\overline{\sigma}}
\newcommand{\boldX}{\mathbf{X}}
\newcommand{\boldW}{\mathbf{W}}
\newcommand{\boldf}{\mathbf{f}}
\newcommand{\dboldX}{\mbox{d}\mathbf{X}}
\newcommand{\dboldW}{\mbox{d}\mathbf{W}}
\newcommand{\dt}{\mbox{d}t}
\newcommand{\dr}{\mbox{d}r}
\newcommand{\dc}{\mbox{d}c}
\newcommand{\dhr}{\mbox{d}\hat{r}}
\newcommand{\dhc}{\mbox{d}\hat{c}}
\newcommand{\picturesAB}[8]{
\centerline{
\hskip #4
\raise #3 \hbox{\raise 0.9mm \hbox{(a)}}
\hskip -8mm
\epsfig{file=#1,height=#3}
\raise #3 \hbox{\raise 0.9mm \hbox{(b)}}
\hskip -8mm
\epsfig{file=#2,height=#3}
\begin{picture}(0,0)
\put(-382,96){#5}
\put(-286,-4){#6}
\put(-195,96){#7}
\put(-110,-4){#8}
\end{picture}
\addtocounter{myfigure}{1}
}}

\newcounter{myfigure}
\renewcommand{\thefigure}{\themyfigure}

\title{Analysis of Brownian Dynamics Simulations of Reversible 
Bimolecular Reactions}

\author{\quad  
Jana Lipkov\'a\thanks{
Charles University, Faculty of Mathematics and Physics,
Sokolovsk\'a 83, 186 75 Prague 8, Czech Republic;
{\it e-mail:  j.lipkova@gmail.com}.}
\quad \and \quad  Konstantinos C. Zygalakis\thanks{University of Oxford, Mathematical Institute, 
24-29 St. Giles', Oxford, OX1 3LB, United Kingdom;
{\it e-mails:  zygalakis@maths.ox.ac.uk; chapman@maths.ox.ac.uk; 
erban@maths.ox.ac.uk}.} \quad
\and \quad  
S. Jonathan Chapman\footnotemark[2]
\and \quad  
Radek Erban\footnotemark[2]
}

\begin{document}

\maketitle

\begin{abstract}
A class of Brownian dynamics algorithms for stochastic reaction-diffusion
models which include reversible bimolecular reactions is presented and 
analyzed. The method is a generalization of the $\lambda$--$\newrho$ model 
for irreversible bimolecular reactions which was introduced in 
\cite{Erban:2009:SMR}. The formulae relating the experimentally measurable 
quantities (reaction rate constants and diffusion constants) with the 
algorithm parameters are derived. The probability of geminate 
recombination is also investigated. 
\end{abstract}


\pagestyle{myheadings}
\thispagestyle{plain}
\markboth{J. LIPKOV\'A, K. ZYGALAKIS, J. CHAPMAN, R. ERBAN}
{BROWNIAN DYNAMICS OF REVERSIBLE BIMOLECULAR REACTIONS}

\section{Introduction}
Brownian dynamics algorithms are used in a number of application
areas, including modelling of ion channels \cite{Corry:2000:TCT},
macromolecules \cite{Larson:1999:BDS}, liquid crystals 
\cite{Siettos:2003:CBD} and biochemical reaction
networks \cite{Lipkow:2005:SDP} to name a few. The main
idea is that some components of the system (e.g. solvent molecules), 
which are of no special interest to a modeller, are not explicitly 
included in the simulation, but contribute to the dynamics of 
Brownian particles collectively as a random force. This reduces the
dimensionality of the problem, making Brownian dynamics
less computationally intensive than the corresponding
molecular dynamics simulations. In a typical scenario, the
position $\boldX_i = [X_i(t),Y_i(t),Z_i(t)]$ of the Brownian 
particle evolves according to the stochastic differential
equation
\begin{equation} 
\dboldX_i 
=
\boldf_i( \boldX_1, \boldX_2, \dots, \boldX_i, \dots) 
\, 
\dt
+
\sqrt{2D_i}
\, \dboldW_i,
\label{sdewithdrift} 
\end{equation}
where $\boldW_i = [W_{i,x},W_{i,y},W_{i,z}]$ is the standard Brownian  
motion, $D_i$ is the diffusion constant and $\boldf_i$ is the deterministic
drift term which depends on the positions of other 
Brownian particles. Depending on the particular application
area, the drift term $\boldf_i$ can take into account both attractive 
(e.g. electrical forces between ions of the opposite charge), 
repulsive (e.g. steric effects, electrical forces between ions) 
and hydrodynamic interactions \cite{Ermak:1978:BDH}. In this paper, 
we focus on algorithms for spatial simulations of biochemical reaction 
networks in molecular biology. In this application area
\cite{Andrews:2004:SSC,Erban:2009:SMR,vanZon:2005:GFR},
it is often postulated that $\boldf_i \equiv 0$,
i.e. the trajectory of each particle is simply given by
\begin{equation} 
\dboldX_i 
=
\sqrt{2D_i}
\, \dboldW_i.
\label{sdediffusion} 
\end{equation}
In \cite{Erban:2009:SMR}, we used this description of molecular
trajectories and analyzed the so called $\lambda$--$\newrho$
stochastic simulation algorithm for modelling irreversible 
bimolecular reactions. Considering three chemical species
$A,$ $B$ and $C$ which are subject to the bimolecular reaction
\begin{equation}
A + B 
\;
\displaystyle\mathop{\displaystyle\longrightarrow}^{k_1}
\; 
C,
\label{irreversiblereaction}
\end{equation}
it is postulated that a molecule of $A$ and a molecule of
$B$ react with the rate $\lambda$ whenever their distance
is smaller than the binding (reaction) radius $\newrho$. This
definition makes use of two parameters $\lambda$ and
$\newrho$ while the irreversible reaction 
(\ref{irreversiblereaction}) is described in terms
of one parameter, the reaction rate $k_1$. Consequently,
there exists a curve in the $\lambda$--$\newrho$ parameter
space which corresponds to the same rate constant $k_1$.
In the limit $\lambda \to \infty$, the model reduces
to the classical Smoluchowski description of diffusion-limited
reactions, namely, two molecules always react whenever
they are closer than the reaction radius $\newrho$
\cite{Smoluchowski:1917:VMT,Berg:1983:RWB}. However,
having two parameters $\lambda$ and $\newrho$, we can
choose the reaction radius $\newrho$ close to the 
molecular radius (which is often larger than the radius
given by the Smoluchowski model 
\cite{Northrup:1992:KPP,Erban:2009:SMR}) and use $k_1$ to 
compute the appropriate value of $\lambda$. In this paper,
we will study extensions of the $\lambda$--$\newrho$ model
to the reaction-diffusion systems which include reversible 
biochemical reactions of the form
\begin{equation}
{\mbox{ \raise 0.4 mm \hbox{$A+B$}}}
\;
\mathop{\stackrel{\displaystyle\longrightarrow}\longleftarrow}^{k_1}_{k_2}
\;
{\mbox{\raise 0.4 mm\hbox{$C.$}}}
\label{reversiblereaction}
\end{equation}
This reaction effectively means two reactions, the forward
reaction (\ref{irreversiblereaction}) which is modelled with the
help of two parameters $\lambda$ and $\newrho$ 
(as studied in \cite{Erban:2009:SMR}) and the backward 
reaction
\begin{equation}
C 
\;
\displaystyle\mathop{\displaystyle\longrightarrow}^{k_2}
\; 
A + B
\label{reversestep}
\end{equation}
which can be also implemented in terms of two parameters:
the rate constant of the dissociation of the complex
$C$ and the unbinding radius $\newsigma$. Since the reaction
(\ref{reversestep}) is of the first-order, the cleavage of
the complex $C$ is a Poisson process with the rate constant
$k_2$, i.e. the rate constant of the dissociation of $C$
is equal to the experimentally measurable quantity $k_2$. 
The second parameter, the unbinding radius $\newsigma$,
is the initial separation of the molecules of $A$ and $B$ 
which are created after a molecule of $C$ dissociates.

Whenever new molecules of $A$ and $B$ are introduced to
the system, we have to initiate their positions. Since
the algorithm considers all molecules as points, 
it would make sense to place them at the position where the 
complex $C$ was just before the reaction (\ref{reversestep})
occurred, i.e. we would put $\newsigma=0.$ However, this choice
of $\newsigma$ can be problematic. For example, in the
Smoluchowski limit $\lambda \to \infty$, if two particles
start next to each other, they must immediately react
again according to the forward step (\ref{irreversiblereaction}).
Andrews and Bray \cite{Andrews:2004:SSC} propose a solution to 
this problem by requiring that  the initial separation of molecules, the
unbinding radius $\bar{\sigma}$,  must be greater than the
binding radius $\bar{\varrho}$. Here, we generalize the concept
of unbinding radius for the $\lambda$--$\newrho$ model introduced 
in \cite{Erban:2009:SMR}. Since $\lambda$ is in general less than 
infinity, we can choose the unbinding radius $\newsigma$
which is less than the binding radius $\newrho$, including
the case $\newsigma=0$. This is investigated in detail 
in Section \ref{sigmasmall}, but we start with the case 
$\newsigma > \newrho$ in Section \ref{sigmalarge}.

\begin{figure}
\centerline{
\hskip 2mm
\raise 2in \hbox{\raise 0.9mm \hbox{(a)}}
\hskip -8mm
\epsfig{file=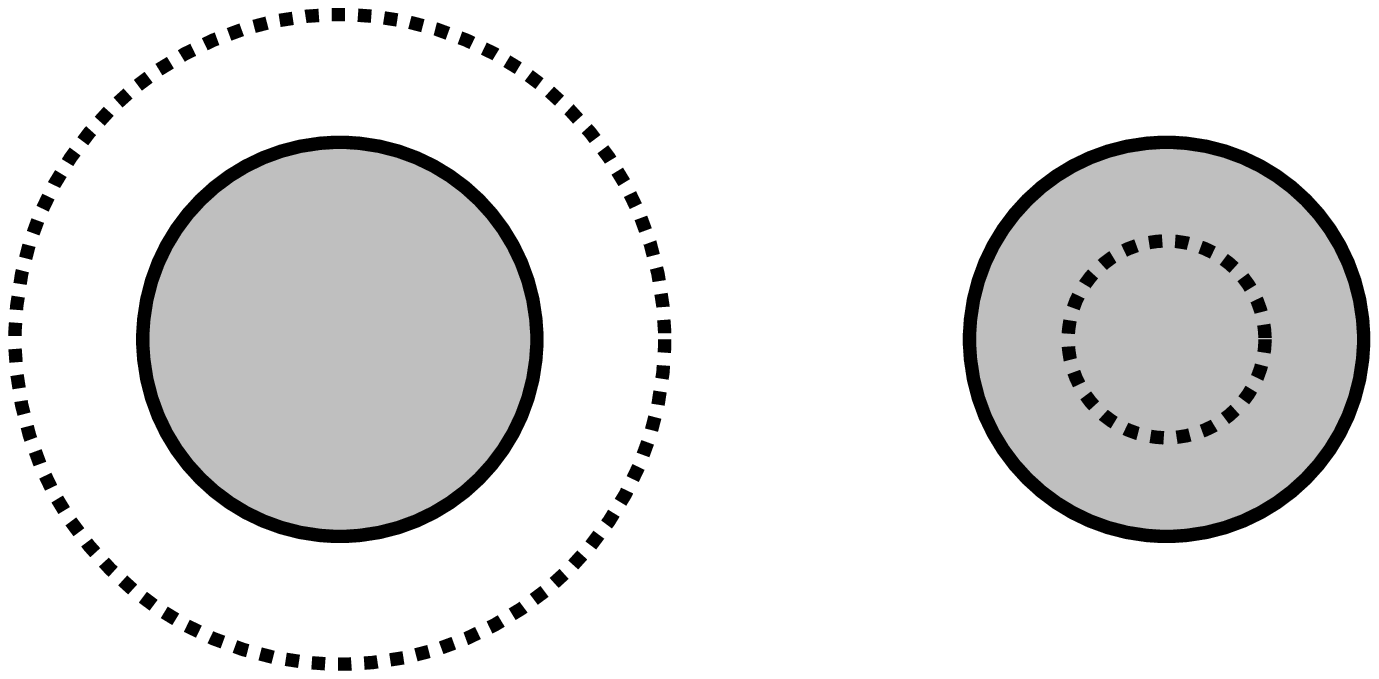,height=2in}
\hskip 1.8mm
\raise 2in \hbox{\raise 0.9mm \hbox{(b)}}
\hskip -8mm
\epsfig{file=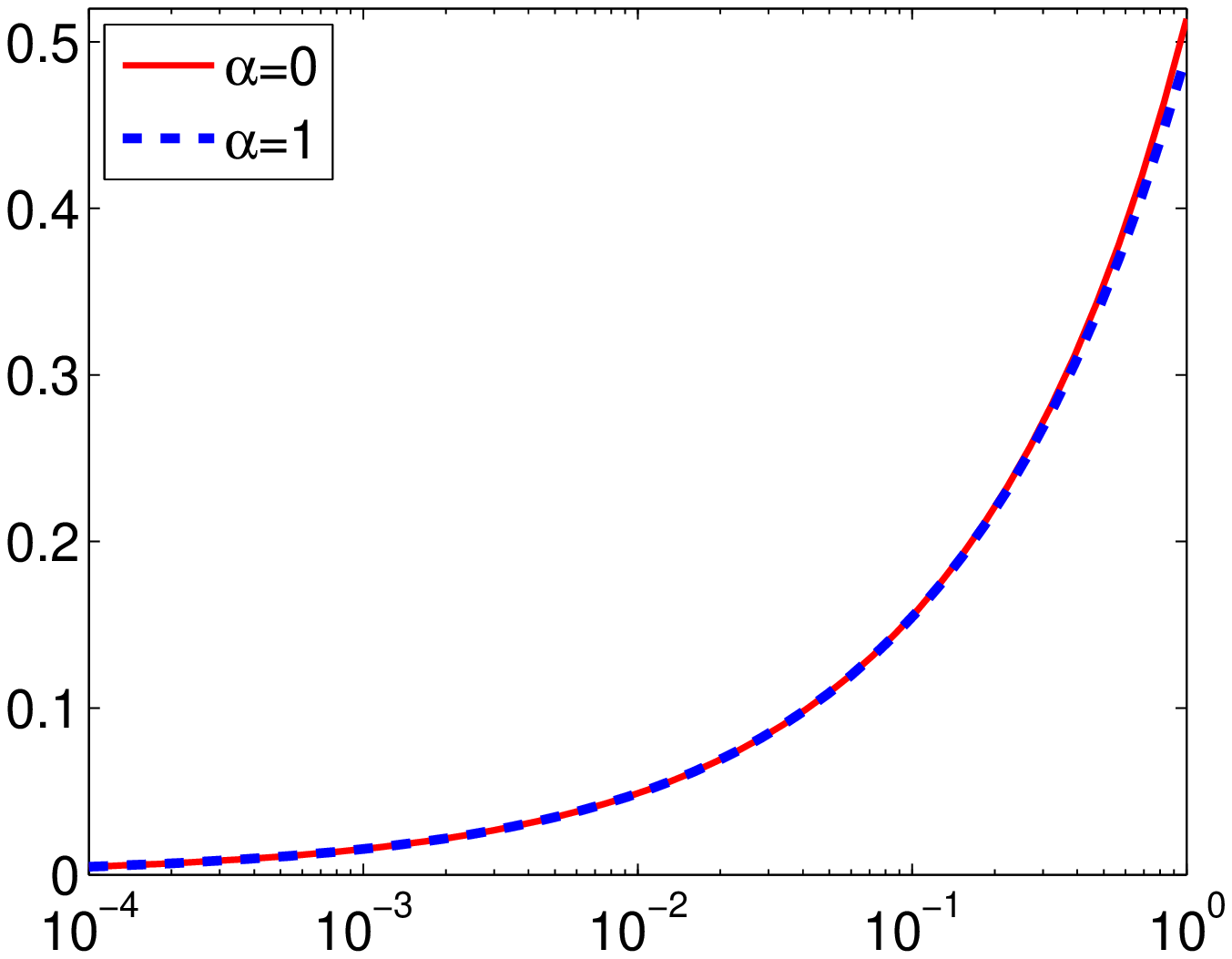,height=2in}
\begin{picture}(0,0)
\put(-302,100){$\overline{\sigma}$}
\put(-260,69){$\overline{\sigma}$}
\put(-315,62){$\overline{\varrho}$}
\put(-229,62){$\overline{\varrho}$}
\put(-350,27){Section \ref{sigmalarge}}
\put(-265,27){Section \ref{sigmasmall}}
\put(-344,128){$\alpha > 1$}
\put(-258,128){$\alpha \le 1$}
\put(-193,92){$\beta$}
\put(-100,-3){$\overline{\kappa}$}
\end{picture}
}
\addtocounter{myfigure}{1}
\caption{(a)
{\it Two cases studied in Sections $\ref{sigmalarge}$ and $\ref{sigmasmall}$.}
(b) {\it Dimensionless parameter
$\beta$ defined by $(\ref{e:dim1})$ as a function of 
$\overline{\kappa}$ for $\alpha=0$ (red solid line) and 
$\alpha=1$ (blue dashed line).}}
\label{P_ratio}
\end{figure}

The algorithm for simulating (\ref{reversiblereaction}) has four parameters:
the binding radius $\newrho$, the unbinding radius $\newsigma$,
the reaction rate $\lambda$ (for the forward step 
(\ref{irreversiblereaction})) and the rate of dissociation of $C$,
but we usually only have two experimentally measurable parameters
$k_1$ and $k_2$. Since $k_2$ is equal to the rate of dissociation
of $C$, the remaining parameters $\lambda$, $\newrho$ and $\newsigma$
will be related to $k_1$. To simplify the derivation of this relation, 
we define the dimensionless parameter $\alpha$ as the
ratio of the unbinding and binding radii, i.e.
\begin{equation}
\alpha = \frac{\newsigma}{\newrho}.
\label{defalpha}
\end{equation}
Two cases are considered separately: $\alpha>1$ and
$\alpha \le 1$, see Figure \ref{P_ratio}(a).
If $\alpha>1$, then the unbinding radius $\newsigma$ is larger
than the binding radius $\newrho$. This situation is investigated 
in Section \ref{sigmalarge}. In Section \ref{sigmasmall}, 
we consider the case $\alpha \le 1$. 
The formula relating $k_1$ with model parameters $\lambda$, 
$\newrho$ and $\newsigma$ is derived
as (\ref{e:nice1}) for $\alpha > 1$ (i.e. for $\newsigma>\newrho$) 
and as (\ref{e:nice2}) for $\alpha \le 1$ (i.e. for
$\newsigma \le \newrho$). 
It is given as one equation for three unknowns 
$\lambda$, $\newrho$ and $\newsigma$.  
In particular, there is a relative freedom in choosing the parameters.
For example, considering that $\newrho$ and $\newsigma$ are given, 
the equations (\ref{e:nice1}) and (\ref{e:nice2}) can be used to
compute the appropriate value of $\lambda$. However, the binding
and unbinding radii are not entirely a choice of a modeller.
This is discussed in Section \ref{gemrecomb}. First,
we would like the binding (reaction) radius to be of a  size similar
to the molecular radius \cite{Erban:2009:SMR}. Second, we sometimes 
want to construct algorithms with a given value of the
probability of geminate recombination
\cite{Andrews:2004:SSC,Andrews:2005:SRL}, which is the probability
that a molecule of $A$ and a molecule of $B$, created from the same
molecule of $C$ by reaction (\ref{reversestep}), react with each other 
according to (\ref{irreversiblereaction}).
In Section \ref{gemrecomb}, we discuss how this extra knowledge
can be used to find optimal values of the parameters of the algorithm.
In particular, we find (equation (\ref{e:nice3})) that the geminate 
recombination probability is proportional to the inverse of the binding 
radius $\newrho$ for the parameter regime relevant to protein-protein
interactions.

The analysis in Sections \ref{sigmalarge}, \ref{sigmasmall}
and \ref{gemrecomb} is done in the limit of (infinitesimally)
small time steps \cite{Erban:2009:SMR}. This provides valuable insights
and a lot of interesting asymptotic behaviour of the algorithm
can be investigated.  However, if we want to implement the $\lambda$--$\newrho$
model on the computer, we have to discretize the stochastic
differential equation (\ref{sdediffusion}) with a finite
time step $\Delta t$ which we want to choose as large
as possible to decrease the computational intensity of the
algorithm. This is studied in Section \ref{secnumer}. The numerical
impementation of the Brownian dynamics algorithm illustrating the 
validity of our analysis is presented in Section \ref{secnumerresults}. 

\section{The case $\alpha > 1$} 
\label{sigmalarge} The $\lambda$--$\newrho$ model of the forward chemical 
reaction (\ref{irreversiblereaction}) states that molecules of $A$ and 
molecules of $B$ diffuse with the diffusion constants $D_A$ and
$D_B$, respectively. If the distance of a molecules of $A$ and a molecule
of $B$ is less than $\newrho$, then the molecules react with the rate
$\lambda$. Considering a frame of reference situated in the molecule
of $B$, we can equivalently describe this process as the random walk 
of a molecule of $A$ which has the diffusion constant $D_{A}+D_{B}$. 
This molecule diffuses to the ball of radius $\newrho$ 
(centered at origin) which removes 
molecules of $A$ with the rate $\lambda$ \cite{Erban:2009:SMR}. In this 
frame of reference, 
the reverse step (\ref{reversestep}) corresponds to the introduction 
of new molecules of $A$ at the distance $\newsigma$ from the origin. 
Let $c(r)$ be the equilibrium concentration of molecules of $A$ 
at distance $r$ from the origin. It is a continuous function 
with continuous derivative which satisfies the following equation:
\begin{eqnarray}
(D_A+D_B) 
\left(
\frac{\mbox{d}^{2}c}{\dr^{2}}+\frac{2}{r}\frac{\dc}{\dr}
\right)
- 
\lambda c &=& 0, \qquad \text{for} \ r\leq \newrho, 
\label{equationalpha1dim1}
\\
(D_A+D_B) 
\left(
\frac{\mbox{d}^{2}c}{\dr^{2}}+\frac{2}{r}\frac{\dc}{\dr}
\right)
+
Q(r-\newsigma) 
&=& 0,\qquad \text{for} \ r\geq \newrho,
\label{equationalpha1dim2}
\end{eqnarray}
where $Q(r-\newsigma)$ is a Dirac-like distribution describing
the creation of molecules at $r = \newsigma$. Let $c_\infty$ be
the concentration of molecules of $A$ in the bulk, i.e.
\begin{equation}
\lim_{r \to \infty} c(r) = c_\infty.
\label{boundaryatinfin}
\end{equation}
To analyze (\ref{equationalpha1dim1})--(\ref{equationalpha1dim2}), we 
define the following dimensionless quantities 
\begin{equation}
\beta=\newrho \, \sqrt{\frac{\lambda}{D_A+D_B}} \, , 
\qquad\;
\overline{\kappa} = \frac{k_1}{\newrho \, (D_A + D_B)} \, ,
\qquad\;
\hat{r}=\frac{r}{\newrho} \, ,
\qquad\;
\hat{c}=\frac{c}{c_\infty} \, ,
\label{e:dim1}
\end{equation}
which means that we scale lengths with $\newrho$ and
times with $\newrho^2 (D_A+D_B)^{-1}$. Substituting (\ref{e:dim1}) 
into (\ref{equationalpha1dim1})--(\ref{equationalpha1dim2}),
we obtain
\begin{eqnarray}
\frac{\mbox{d}^{2}\hat{c}}{\dhr^2}
+
\frac{2}{\hat{r}}\frac{\dhc}{\dhr}-\beta^{2} \, \hat{c} 
&=& 0, \qquad \text{for} \ \hat{r}\leq 1, 
\label{equationalpha1NONdim1}\\
\frac{\mbox{d}^{2}\hat{c}}{\dhr^2}
+\frac{2}{\hat{r}}\frac{\dhc}{\dhr}
+\omega \, \delta(\hat{r}-\alpha) 
&=& 0,\qquad \text{for} \ \hat{r}\geq 1,
\label{equationalpha1NONdim2}
\end{eqnarray}
where $\omega$ is the rate of creation of molecules at 
$\hat{r} = \alpha$. To determine $\omega$, let us note
that the average number of molecules of $A$ produced by
the reverse step (\ref{reversestep}) is (at equilibrium)
equal to the average number of molecules of $A$ destroyed
by the forward reaction (\ref{irreversiblereaction}),
i.e. the equilibrium flux through the sphere of radius
1 is equal to $4 \pi \alpha^2 \omega$. This implies
\begin{equation}
4 \pi \alpha^2 \omega
=
4 \pi \, \frac{\dhc}{\dhr} \Big|_{\hat{r}=1}.
\label{condonomega}
\end{equation}
The right hand side of (\ref{condonomega}) is also equal
to the dimensionless rate constant $\overline{\kappa}$
of the forward reaction (\ref{irreversiblereaction}).
Consequently, we get $4 \pi \alpha^2 \omega = \overline{\kappa}$.
Substituting $\overline{\kappa}/(4 \pi \alpha^2)$ for $\omega$, 
the general solution 
of (\ref{equationalpha1NONdim1})--(\ref{equationalpha1NONdim2}) 
can be written in the following form
\begin{eqnarray}
\hat{c}(\hat{r}) &=& 
\frac{a_1}{\hat{r}} \, e^{\beta \hat{r}}
+
\frac{a_2}{\hat{r}} \, e^{-\beta \hat{r}},
\qquad \qquad \qquad \qquad \qquad \; \; \,
\text{for} \ \hat{r}\leq 1, 
\label{pp1}
\\
\hat{c}(\hat{r}) 
&=& 
a_3
-
\frac{a_4}{\hat{r}}
-
\frac{\overline{\kappa} \, H(\hat{r}-\alpha)(\hat{r}-\alpha)}{4 \pi 
\hat{r}\alpha}, 
\qquad \qquad \quad \ \text{for} \ \hat{r}\geq 1, 
\label{pp2}
\end{eqnarray}
where $H$ denotes the Heaviside step function and $a_1,$
$a_2,$ $a_3,$ $a_4$ are real constants to be determined.  
The boundary condition (\ref{boundaryatinfin}) at infinity 
in the dimensionless variables read as follows
\begin{equation}
\lim_{\hat{r} \rightarrow \infty} \hat{c}(\hat{r})=1.
\label{bouninfhat}
\end{equation}
Using this condition, the continuity of $\hat{c}$ at 
the origin, and the continuity of $\hat{c}$ and its 
derivative at $\hat{r} = 1$, we determine the 
constants $a_1,$ $a_2,$ $a_3$ and $a_4$ in 
(\ref{pp1})--(\ref{pp2}). We obtain
\begin{eqnarray*}
\hat{c}(\hat{r}) 
&=& 
\frac{4 \pi \alpha + \overline{\kappa}}
{4 \pi \alpha \, \beta \, \cosh \beta} 
\, \frac{\sinh{\beta \hat{r}}}{\hat{r}}, 
\qquad\qquad\qquad\qquad\qquad 
\qquad\qquad\qquad\quad 
\text{for} \ \hat{r}\leq 1, \\
\hat{c}(\hat{r}) 
&=& 
\frac{4 \pi \alpha + \overline{\kappa}}{4 \pi \alpha}
\left(
1 
-
\frac{1}{\hat{r}}
+
\frac{\tanh \beta}{\beta \, \hat{r}}
\right)
-
\frac{\overline{\kappa} \, H(\hat{r}-\alpha)(\hat{r}-\alpha)}{4 \pi 
\hat{r}\alpha},
\qquad \qquad \ \ \text{for} \ \hat{r}\geq 1.
\end{eqnarray*}
Substituting $\hat{c}$ into (\ref{condonomega}) where
$4 \pi \alpha^2 \omega = \overline{\kappa}$, we obtain
\begin{equation} \label{e:nice1_dim_less}
\overline{\kappa}
= 
\frac{4 \pi \alpha \, (\beta-\tanh{\beta})}
{\beta \, \alpha - \beta +\tanh{\beta}}  
\end{equation}
which is the desired relation between the measurable quantities
and the model parameters. Using
(\ref{defalpha}) and (\ref{e:dim1}), the condition 
\eqref{e:nice1_dim_less} can be equivalently expressed in terms 
of the measurable rate constant $k_1$ and diffusion constants
$D_A$, $D_B$, and the model parameters (binding radius 
$\newrho$, unbinding radius $\newsigma$ and the rate 
$\lambda$) as follows
\begin{equation}
k_{1}
= 
\frac{4\pi\newsigma(D_{A}+D_{B})
\left(\newrho\sqrt{\frac{\lambda}{D_{A}+D_{B}}}
-
\tanh\left(\newrho\sqrt{\frac{\lambda}{D_{A}+D_{B}}}\right) \right)}
{\newsigma\sqrt{\frac{\lambda}{D_{A}+D_{B}}}
-
\newrho\sqrt{\frac{\lambda}{D_{A}+D_{B}}}
+
\tanh\left(\newrho\sqrt{\frac{\lambda}{D_{A}+D_{B}}}\right)}. 
\label{e:nice1} 
\end{equation}

\noindent
{\bf Remark:} If we take the limit of $\alpha \to \infty$ in 
(\ref{e:nice1_dim_less}), we obtain
\begin{equation} 
\lim_{\alpha \to \infty} \overline{\kappa}
= 
4 \pi (1 - \beta^{-1} \tanh{\beta}).  
\label{limitalphainfinity}
\end{equation}  
This is exactly the same expression as in \cite{Erban:2009:SMR}  
for the original $\lambda$--$\newrho$ model, which describes 
only the bimolecular reaction (\ref{irreversiblereaction}).
However, this should not be a surprise, since by taking the limit 
$\alpha \rightarrow \infty$, we effectively remove the  
reverse reaction (\ref{reversestep}) from the system. 
Passing to the limit $\beta \to \infty$ in
(\ref{limitalphainfinity}), we obtain the relation
\begin{equation}\label{e:Smolu}
k_1 = 4 \pi \rho_{s} (D_A + D_B)
\end{equation} 
where $\rho_{s}$ is the radius in the Smoluchowski model of diffusion-limited reactions \cite{Erban:2009:SMR,Smoluchowski:1917:VMT}.

\section{The case $\alpha \le 1$} 
\label{sigmasmall} If $\newsigma\leq \newrho$, then the 
equilibrium equations
(\ref{equationalpha1NONdim1})--(\ref{equationalpha1NONdim2})
together with the boundary condition (\ref{bouninfhat})
at infinity have to be replaced by one equation
\begin{equation}
\frac{\mbox{d}^{2}\hat{c}}{\dhr^2}
+
\frac{2}{\hat{r}}\frac{\dhc}{\dhr}-\beta^{2} \, \hat{c}
+ \frac{\overline{\kappa} \, \delta(\hat{r}-\alpha)}{4 \pi \alpha^2}  
= 0, \qquad \text{for} \ \hat{r}\leq 1, 
\label{equationalphasmallNON}
\end{equation}
with the boundary condition $\hat{c}(1) = 1.$ This takes into
account the fact that there is no diffusive flux for $\hat{r} > 1$,
i.e. $\hat{c}(\hat{r}) = 1$ for $\hat{r} > 1$. 
The general solution of the second-order ordinary
differential equation (\ref{equationalphasmallNON})
is given by
\begin{equation*}
\hat{c}(\hat{r}) 
=
\frac{a_1}{\hat{r}} \, e^{\beta \hat{r}}
+
\frac{a_2}{\hat{r}} \, e^{-\beta \hat{r}}
-
\frac{\overline{\kappa} \, 
H(\hat{r}-\alpha) \, \sinh (\beta \hat{r} -  \beta \alpha)}
{4 \pi \alpha \beta \, \hat{r}},
\end{equation*}
where $a_1$ and $a_2$ are real constants which
are determined by the boundary condition $\hat{c}(1) = 1$
and the continuity of $\hat{c}$ at 
the origin. We obtain
\begin{equation}
\hat{c}(\hat{r}) 
=
\frac{4 \pi \alpha \beta + \overline{\kappa} \, \sinh (\beta - \beta \alpha)}
{4 \pi \alpha \, \beta \, \sinh \beta} 
\, 
\frac{\sinh{\beta \hat{r}}}{\hat{r}}
-
\frac{\overline{\kappa} \, 
H(\hat{r}-\alpha) \, \sinh (\beta \hat{r} - \beta  \alpha)}
{4 \pi \alpha \beta \, \hat{r}}.
\label{solutionsmallalpha}
\end{equation}
Since there is no diffusive flux at $\hat{r} = 1$ at equilibrium, we have 
$$
\frac{\dhc}{\dhr}(1) = 0.
$$
Evaluating this condition for (\ref{solutionsmallalpha}),
we get
\begin{equation} 
\label{e:nice2_dim_less}
\overline{\kappa}
=
\frac{4\pi\alpha \, (\beta-\tanh \beta)}
{\cosh(\beta -  \beta \alpha) \tanh \beta - \sinh (\beta -  \beta \alpha)} 
\end{equation}
which can be expressed in terms of the experimentally
measurable quantities $k_1$, $D_A$ and $D_B$, and 
the model parameters $\newrho$, $\newsigma$ and $\lambda$
as 
\begin{equation} \label{e:nice2}
k_{1}
=
\frac{4\pi\newsigma (D_{A}+D_{B})
\left(\newrho\sqrt{\frac{\lambda}{D_{A}+D_{B}}} 
-
\tanh\left(\newrho\sqrt{\frac{\lambda}{D_{A}+D_{B}}}\right)\right)}
{ \cosh\left((\newrho-\newsigma)\sqrt{\frac{\lambda}{D_{A}+D_{B}}} \right)
\tanh{\left(\newrho\sqrt{\frac{\lambda}{D_{A}+D_{B}}} \right)}
- \sinh\left((\newrho-\newsigma)\sqrt{\frac{\lambda}{D_{A}+D_{B}}} \right)}.
\end{equation}

\subsection{Asymptotic behaviour}
\label{secasymbeh1}
Let us consider that the binding radius $\newrho$ is fixed.
Since $k_1$, $D_A$ and $D_B$ are typically given by 
experiments, the dimensionless parameter $\overline{\kappa}$
is a fixed nonnegative constant. Taking the limit 
$\alpha \to 0$ in (\ref{e:nice2_dim_less}), we obtain
\begin{equation} \label{e:limit2_dim_less}
\lim_{\alpha \to 0} \overline{\kappa}
=
4 \pi (\cosh{\beta} - \beta^{-1} \sinh{\beta}). 
\end{equation}
Since the left-hand side is a nonnegative constant
and the right-hand side an increasing function of $\beta$, 
we can solve (\ref{e:nice2_dim_less}) for $\beta.$ We denote
the unique solution of (\ref{e:nice2_dim_less}) as 
$\beta_{c}$. We have $\beta_{c} > 0$ because the right hand 
side of equation \eqref{e:limit2_dim_less} approaches zero 
in the limit $\beta \to 0$.

Considering typical values of the diffusion and reaction
rate constants for proteins, namely 
$D_{A}=D_{B}=10^{-5}$ cm$^{2}$ s$^{-1}$, 
$\newrho=2$ nm and $k_{1}=10^{6}$ M$^{-1}$,
we find that $\overline{\kappa} \simeq 4.17 \times 10^{-4}$ 
and $\beta_{c} \simeq 10^{-2}$, i.e. both 
$\overline{\kappa}$ and $\beta_{c}$ are small parameters.
Considering small $\beta$ and $\alpha$ of order
1, the leading order term in the expansion of 
(\ref{e:nice2_dim_less}) is $4 \pi \beta^2 / 3$ which
is independent of $\alpha$. Consequently, we observe that 
\eqref{e:limit2_dim_less} is actually a good approximation 
of (\ref{e:nice2_dim_less}) even for $\alpha$ of order 1. 
This point is illustrated in Figure \ref{P_ratio}(b), where we 
plot $\beta$ defined in \eqref{e:dim1} as a function of 
$\overline{\kappa}$ for $\alpha=0$ and $\alpha=1$. As we can see, 
it is only when $\overline{\kappa}$ becomes of order 1 that the 
rate $\beta$ calculated with \eqref{e:nice2_dim_less} slightly
differs from 
the one calculated using \eqref{e:limit2_dim_less}. This implies 
that there exist a realistic parameter regime for 
$\overline{\kappa}$ 
for which the parameter $\alpha$ is not influencing the value 
of the removal rate $\beta$, and $\alpha$ can thus be set to 
0 or 1. 
Morever, this implies for this particular parameter range of 
$\overline{\kappa}$, we can completely drop the concept of 
the unbinding radius $\newsigma$.

\section{Geminate recombination}
\label{gemrecomb}
In Sections \ref{sigmalarge} and \ref{sigmasmall},
we derived formulae (\ref{e:nice1}) and (\ref{e:nice2})
relating the algorithm parameters with the experimentally 
measurable quantities. In both cases $\alpha>1$ and
$\alpha \le 1$, we have one equation for three unknowns
$\newrho$, $\newsigma$ and $\lambda$. The binding radius 
$\newrho$ describes the range of interaction between
molecules. Postulating that $\newrho$ is comparable
to the experimentally measurable molecular radius,
we are left with two unknows $\newsigma$ and $\lambda$
related by one condition (\ref{e:nice1}) (resp. (\ref{e:nice2})).
Using the dimensionless parameters (\ref{e:dim1}),
we can also formulate it as one equation (\ref{e:nice1_dim_less})
(resp. (\ref{e:nice2_dim_less})) for two unknowns
$\alpha$ and $\beta$. In particular, different choices 
of these parameters lead to the same reaction rates. 
If we want to uniquely specify $\alpha$ and $\beta$,
we will need an extra equation. In this section, we show 
that different pairs of $\alpha$ and $\beta$ 
(which lead to the same reaction rates) correspond to 
different probability of geminate recombination
(which is properly defined in the next paragraph).
This observation can be used to find the missing
relation between $\alpha$ and $\beta$. 

When a molecule of $C$ dissociates, one molecule of $A$ and 
one molecule of $B$ are introduced to the system. They
can have two possible fates. Either, they react again
to form the same complex $C$, or they diffuse away from 
each other. The first case is called {\it geminate recombination}
\cite{Andrews:2005:SRL,Andrews:2004:SSC}.
We denote by $\phi$ the probability of geminate recombination,
i.e. the probability that the newly born pair of $A$ and $B$
reacts again. To derive a formula relating $\phi$, $\alpha$ 
and $\beta$, we denote by $p(\hat{r})$ the probability that a 
molecule of $A$, which is introduced in distance $\hat{r}$ from 
a molecule of $B$, will react with $B$ before escaping 
to infinity.  The probability $p(\hat{r})$ is a continuous 
function with continuous derivative satisfying the equations   
\begin{eqnarray}
\frac{d^{2} p}{d \hat{r}^{2}}+\frac{2}{\hat{r}}\frac{dp}{d\hat{r}}
 &=&  \beta^{2} (p-1), \qquad \qquad \;\,\text{for} \ \hat{r}\leq 1, 
\label{reactionprob1}\\ 
\frac{d^{2} p}{d \hat{r}^{2}}+\frac{2}{\hat{r}}\frac{dp}{d\hat{r}}
&=& 0, 
\label{reactionprob2}
\qquad\qquad  \qquad \qquad \text{for} \ \hat{r}\geq 1,
\end{eqnarray}
and the boundary condition
\begin{equation}
\lim_{\hat{r}\to \infty} p(\hat{r}) = 0. 
\label{bouninfprobesc}
\end{equation}
Solving (\ref{reactionprob1})--(\ref{bouninfprobesc}), we get
\begin{eqnarray*}
p(\hat{r}) &=& 
1
- 
\frac{\sinh (\hat{r} \beta)}{\hat{r} \, \beta \cosh \beta}
\, ,
\qquad\qquad\qquad\;\;\,  \text{for} \ \hat{r}\leq 1,
\\
p(\hat{r}) &=& 
\frac{\beta-\tanh{\beta}}{\hat{r} \, \beta}\, ,
\qquad\qquad\qquad \qquad \text{for} \ \hat{r}\geq 1.
\end{eqnarray*}
Whenever the reverse reaction (\ref{reversestep}) takes place,
the initial separation of molecules of $A$ and $B$ 
is equal to $\alpha$ (in dimensionless variables). 
Consequently, the probability $\phi$ of geminate 
recombination is given as $\phi=p(\alpha)$, i.e.

\begin{eqnarray}  
\phi &=& 
1
- 
\frac{\sinh (\alpha \beta)}{\alpha \beta \cosh \beta}\,,
\qquad\qquad\quad  \text{for} \ \alpha \leq 1,
\label{phi1}
\\
\phi &=& 
\frac{\beta-\tanh{\beta}}{\alpha \beta},
\qquad\qquad\qquad\;\, \text{for} \ \alpha \geq 1.
\label{phi2}
\end{eqnarray}

If a modeller wants to design an algorithm with a given
value of the probability $\phi$ of geminate recombination,
then equations \eqref{phi1}, \eqref{phi2}  will give the second 
condition relating the parameters $\alpha$
and $\beta$. The first one is (\ref{e:nice1_dim_less})
(resp. (\ref{e:nice2_dim_less})).

\subsection{Asymptotic behaviour}
\begin{figure}
\picturesAB{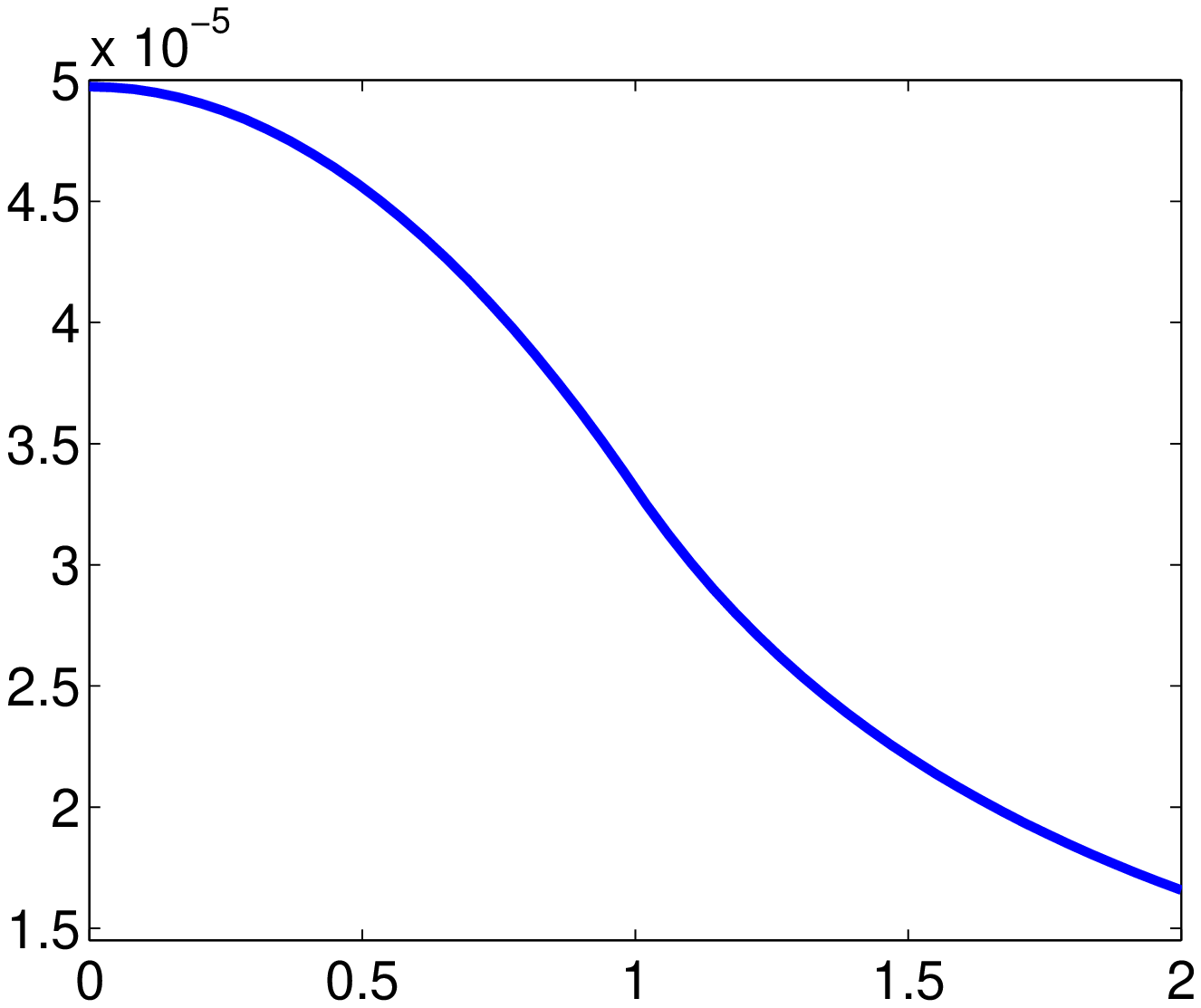}{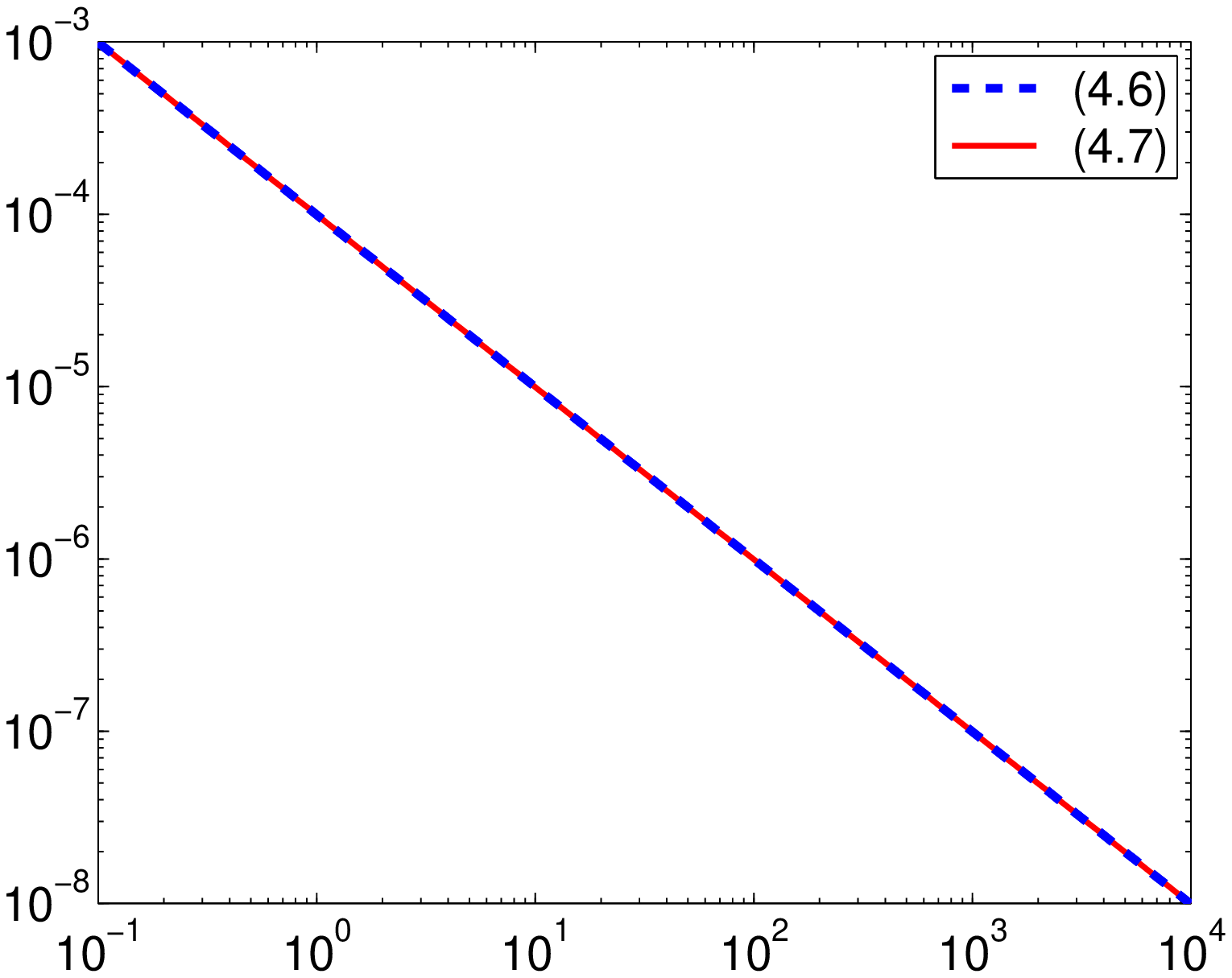}
{2in}{6 mm}{$\phi$}{$\alpha$}{$\phi$}{$\newrho$ [nm]}
\caption{{\rm (a)} {\it  Geminate recombination probability $\phi$ 
as function of $\alpha$. We use $D_{A}=D_{B}=10^{-5}$ cm$^{2}$ s$^{-1}$, 
$\newrho=2$ nm and $k_{1}=10^{6}$ {\rm M}$^{-1}$.} 
{\rm (b)} {\it Comparison of the geminate 
recombination probability $\phi$ calculated by $\eqref{e:sigma_zero}$ 
and $\eqref{e:nice3}$. We use $D_{A}=D_{B}=10^{-5}$ cm$^{2}$ s$^{-1}$ 
and $k_{1}=10^{6}$ {\rm M}$^{-1}$.}} 
\label{fig:geminate1}
\end{figure}

As we observed in Section \ref{secasymbeh1}, realistic
parameters for protein-protein interactions lead to 
a small value of the dimensionless parameter $\beta$. In particular,
the second condition relating $\alpha$ and $\beta$ is not needed
because different values of $\alpha$ lead to the same results. 
Considering the same parameters as in Section \ref{secasymbeh1},
we plot the geminate recombination probability $\phi$ as 
a function of the dimensionless ratio $\alpha$ in
Figure \ref{fig:geminate1}(a). To compute this plot, we 
use \eqref{e:nice1} or \eqref{e:nice2} to calculate $\beta$
for a given value of $\alpha$. Then we calculate $\phi$ 
using \eqref{phi1},\eqref{phi2}. In Figure \ref{fig:geminate1}(a),
we observe that the probability $\phi$ of geminate recombination
is close to zero for all values of $\alpha$. 
\noindent
If $\alpha=0$, then \eqref{phi1} implies
\begin{equation} \label{e:sigma_zero}
\phi=1-\frac{1}{\cosh{\beta_{c}}} \, , 
\end{equation}
where $\beta_{c}$ satisfies \eqref{e:limit2_dim_less}. Since  
$\beta_{c} \ll 1$, equations  \eqref{e:sigma_zero} and 
\eqref{e:limit2_dim_less} give
\[
\phi  \approx \frac{1}{2}\beta^{2}_{c}, 
\qquad \qquad \text{and} \qquad \qquad \overline{\kappa}
\approx \frac{4 \pi \beta^{2}_{c}}{3}.
\]
Combining these two equations  we obtain 
$\phi=3\overline{\kappa}/(8\pi). 
$
Substituting (\ref{e:dim1}) for $\overline{\kappa}$, we get
\begin{equation} \label{e:nice3}
\phi = \frac{3\rho_{s}}{2\newrho} 
\end{equation}
where $\rho_{s}$ is the reaction radius corresponding to 
the Smoluchowski model given by \eqref{e:Smolu}.
In Figure \ref{fig:geminate1}(b), we plot the geminate 
recombination probability $\phi$ as a function of 
$\newrho$ for $\alpha = 0$. We use the same values of
$D_{A},D_{B}$ and $k_{1}$ as in Figure \ref{fig:geminate1}(a)
and we vary $\newrho$ from 1\AA{} (0.1 nm) to thousands
of nanometres. We observe that the formula \eqref{e:sigma_zero}
(together with \eqref{e:limit2_dim_less})
gives the same geminate recombination probability $\phi$ 
as the approximation \eqref{e:nice3}. Finally, let us note
that by taking the limit $\beta \to \infty$ 
in (\ref{phi2}), we obtain $\phi = \alpha^{-1} 
= \newrho/\newsigma,$ which is the expression for the geminate 
recombination probability used in \cite{Andrews:2004:SSC}.

\section{Stochastic simulation algorithm for large time steps}
\label{secnumer}
To implement $\lambda$-$\newrho$ model on a computer, 
we have to discretize \eqref{sdediffusion} using a 
finite time step $\Delta t$. Using the Euler-Maruyama 
method \cite{Platen:1999:INM,Erban:2007:PGS}, the position 
$[X_i(t+\Delta t),Y_i(t+\Delta t),Z_i(t+\Delta t)]$ 
of the $i$-th molecule at time $t+\Delta t$ is computed
from its position $[X_i(t),Y_i(t),Z_i(t)]$ at time $t$ by 
\begin{eqnarray}
X_i(t+\Delta t) &=& X_i(t)+\sqrt{2D_i\Delta t} \, \xi_{x}, 
\nonumber 
\\
Y_i(t+\Delta t) &=& Y_i(t)+\sqrt{2D_i\Delta t} \, \xi_{y}, 
\label{e:solve_SDE1}
\\
Z_i(t+\Delta t) &=& Z_i(t)+\sqrt{2D_i\Delta t} \, \xi_{z}, 
\nonumber
\end{eqnarray}
where $\xi_{x},\xi_{y},\xi_{z}$ are random numbers which are 
sampled from the normal distribution with zero mean and 
unit variance. If $\Delta t$ is ``very small", then the computer
implementation of the reversible reaction (\ref{reversiblereaction})
is straightforward. We use (\ref{e:solve_SDE1}) to update the 
position of every molecule using $D_i=D_{A}$ for molecules of $A$,
$D_i=D_{B}$ for molecules of $B$ and $D_i=D_{C}$ for molecules of $C$.
Whenever the distance of a molecule of $A$ from a molecule of $B$ is 
less than the reaction radius $\newrho$, the molecules react
according to the forward reaction (\ref{irreversiblereaction})
with probability $P_{\lambda} = \lambda \, \Delta t$.
The probability of the reverse reaction (\ref{reversestep})
during one time step is equal to $k_{2} \, \Delta t$. If
the complex $C$ dissociates, then we introduce one molecule of $A$ 
and one molecule of $B$ in a distance $\newsigma$ apart. 

This computer implementation of the reversible reaction 
(\ref{reversiblereaction}) will only work if the time
step $\Delta t$ is chosen so small that 
$P_{\lambda} = \lambda \, \Delta t \ll 1$,
$k_{2} \, \Delta t \ll 1$ and $\gamma \ll 1$,
where $\gamma$ is given by
\begin{equation}
\gamma = \frac{\sqrt{2(D_{A}+D_{B})\Delta t}}{\newrho}, 
\label{defgamma}
\end{equation}
i.e. $\gamma$ is the ratio of the average step size 
in one coordinate during one time step over the reaction 
radius $\newrho$. 
In this section, we show how the restrictions on the time 
step $\Delta t$ can be removed. First of all, the probability 
that the complex $C$ dissociates during the time interval 
$(t,t+\Delta t)$ is equal to $1 - \exp( - k_{2} \, \Delta t)$, 
i.e. the reverse reaction (\ref{reversestep}) is easy to
implement for arbitrary time step $\Delta t$. We simply use
$1 - \exp( - k_{2} \, \Delta t)$ instead of $k_{2} \, \Delta t$
as the probability of dissociation of $C$ during one time
step. To relax the restrictions $\gamma \ll 1$
and $P_{\lambda} = \lambda \, \Delta t \ll 1$,
we slightly reformulate the algorithm \cite{Erban:2009:SMR}.
As before, it will make use of three parameters: the reaction 
radius $\newrho$, the unbinding radius $\newsigma$ and the 
reaction probability $P_{\lambda}$ of the 
forward reaction (\ref{irreversiblereaction}). 
We postulate that a molecule of $A$ and a molecule of $B$ 
(which are closer than the reaction radius $\newrho$) react 
with probability $P_{\lambda} \in (0,1]$ during the 
next time step. Therefore, the computer implementation 
of the reversible reaction (\ref{reversiblereaction}) will 
make use of the following three steps:

\smallskip

{
\leftskip 8mm

\parindent -5mm
{\bf [i]}\, 
If the distance of a molecule of $A$ from a molecule of $B$ 
(at time $t$) is less than the reaction radius $\newrho$, then
generate a random number $r_1$ uniformly distributed in (0,1). 
If $r_1 < P_\lambda$, then the forward reaction (\ref{irreversiblereaction}) 
occurs, i.e. the molecules of $A$ and $B$ are removed from the system
and a new molecule of $C$ is created.

\parindent -6mm
{\bf [ii]}\, 
For each molecule of $C$, generate a random number $r_2$ uniformly 
distributed in (0,1). If $r_2 < 1 - \exp( - k_{2} \, \Delta t)$, then
the reverse reaction (\ref{reversestep}) takes place, i.e.
the complex $C$ dissociates, and one molecule of $A$ 
and one molecule of $B$ are introduced a distance $\newsigma$ apart. 

\parindent -7mm
{\bf [iii]}\, 
Use (\ref{e:solve_SDE1}) to update the position of every molecule.

\par}

\smallskip

\noindent
The steps [i]--[iii] are repeated during every time step. In order
to use this algorithm, we need to find equations relating  parameters 
$\newrho$, $\newsigma$ and $P_{\lambda}$ with the experimentally
measurable quantities. If $\Delta t$ is small, then one condition
is given as \eqref{e:nice1} (resp. \eqref{e:nice2}) where
$P_\lambda = \lambda \, \Delta t \ll 1.$ However, if $P_\lambda$
is close to 1, we have to modify the derivation of these conditions,
replacing partial differential equations 
(\ref{equationalpha1dim1})--(\ref{equationalpha1dim2})
by suitable integral equations \cite{Erban:2009:SMR,Erban:2007:RBC}. 

First of all, the conditions depend on the ordering of steps [i]--[iii], 
i.e. on the ordering of subroutines of the algorithm. Consider
the case $P_\lambda = 1$ and $\alpha = \newsigma/\newrho < 1$. 
If we ordered the subroutines as [ii], [i] and [iii], then each 
dissociation of a complex $C$ in step [ii] would introduce two new molecules
of $A$ and $B$ which are a distance $\newsigma$ apart. Since
[ii] would be immediately followed by [i], the new molecules would have 
to react again because $P_\lambda = 1$ and their separation is less 
than $\newrho$. In particular, there would be no chance to correctly 
implement this model for $P_\lambda = 1$ and 
$\alpha = \newsigma/\newrho < 1$. On the other hand, if we order the 
subroutines as [i], [ii] and [iii], then the dissociation of $C$ is 
followed by diffusion of molecules, i.e. the new molecules of $A$ 
and $B$ can diffuse away of each other. 
In the rest of this paper, we assume that the subroutines are
ordered as [i], [ii] and [iii] during each time step.

\begin{figure} 
\centerline{
\hskip 6mm
\raise 2in \hbox{\raise 0.9mm \hbox{(a)}}
\hskip -8mm
\epsfig{file=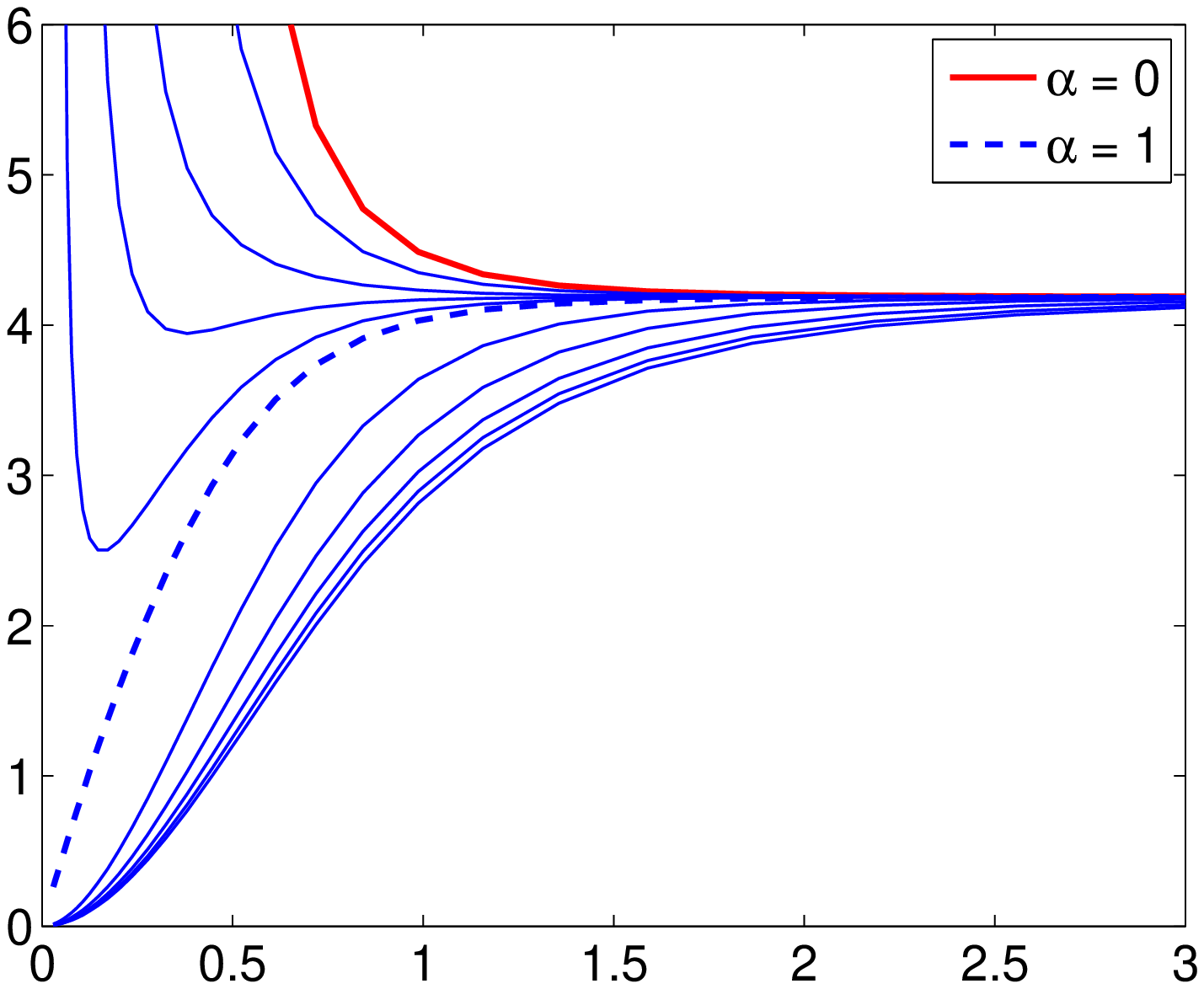,height=2 in}
\raise 2in \hbox{\raise 0.9mm \hbox{(b)}}
\hskip -8mm
\epsfig{file=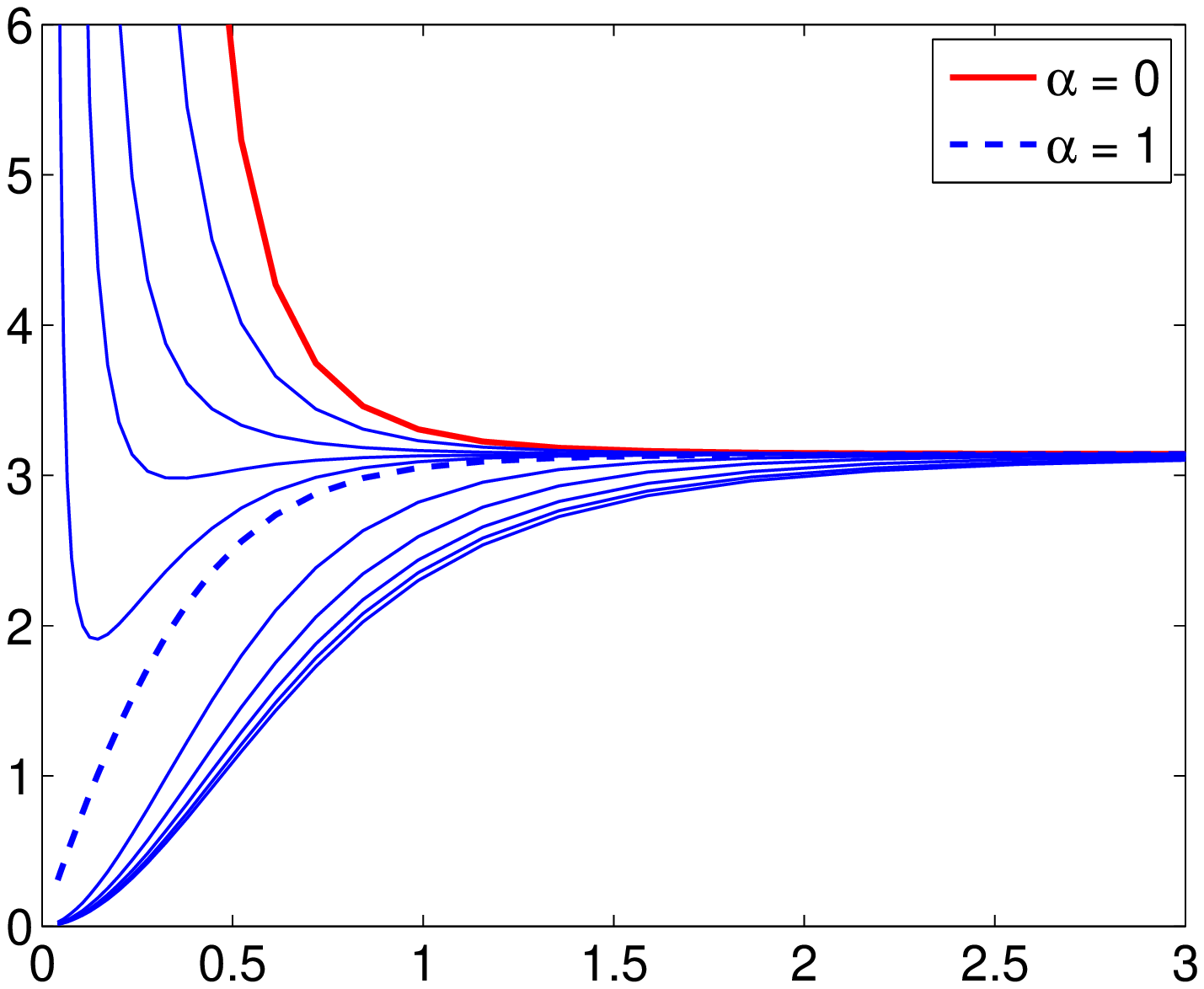,height=2 in}
\begin{picture}(0,0)
\put(-376,83){$\kappa$}
\put(-288,0){$\gamma$}
\put(-186,83){$\kappa$}
\put(-98,0){$\gamma$}
\end{picture}
}
\centerline{
\hskip 6mm
\raise 2in \hbox{\raise 0.9mm \hbox{(c)}}
\hskip -8mm
\epsfig{file=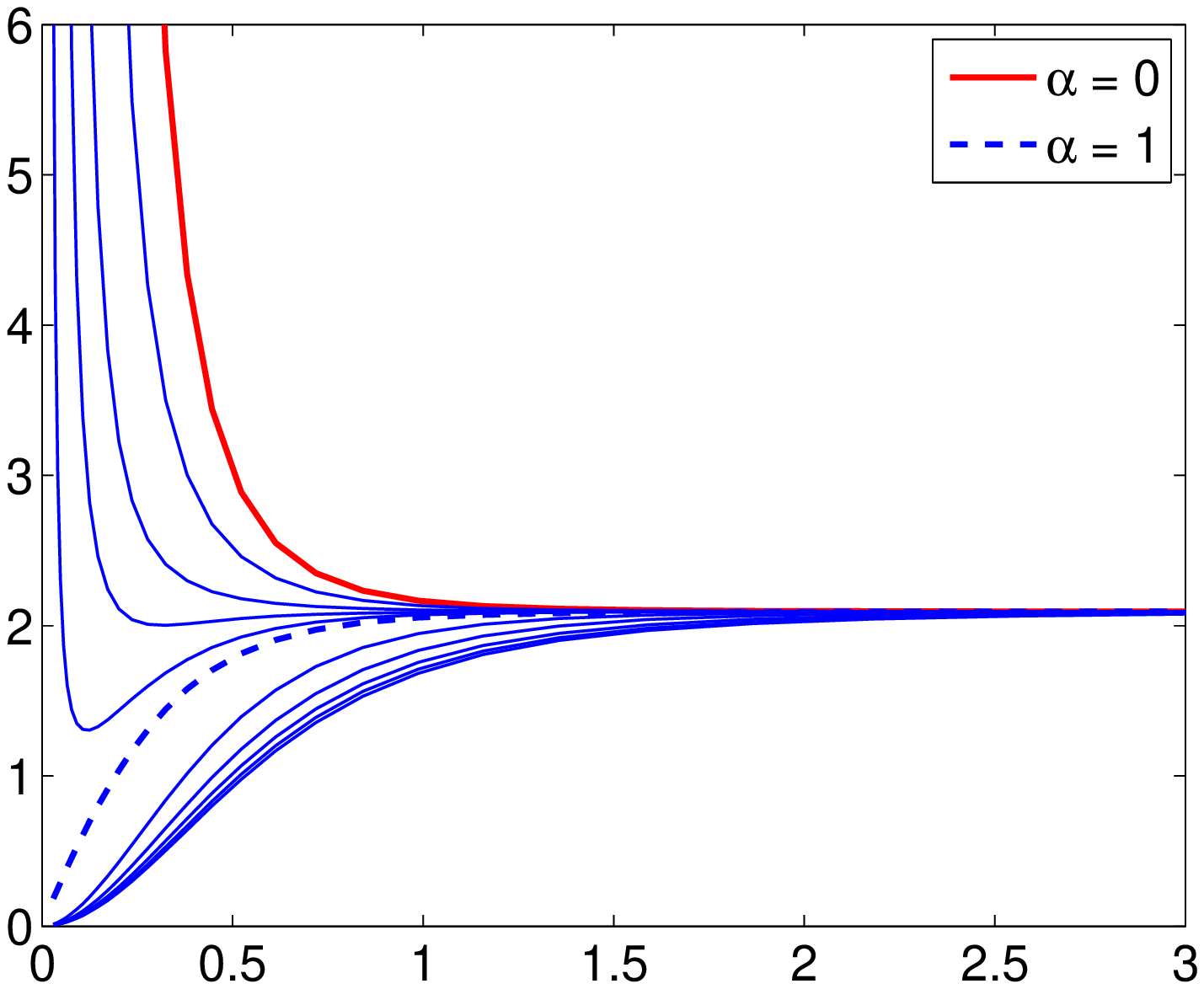,height=2 in}
\raise 2in \hbox{\raise 0.9mm \hbox{(d)}}
\hskip -8mm
\epsfig{file=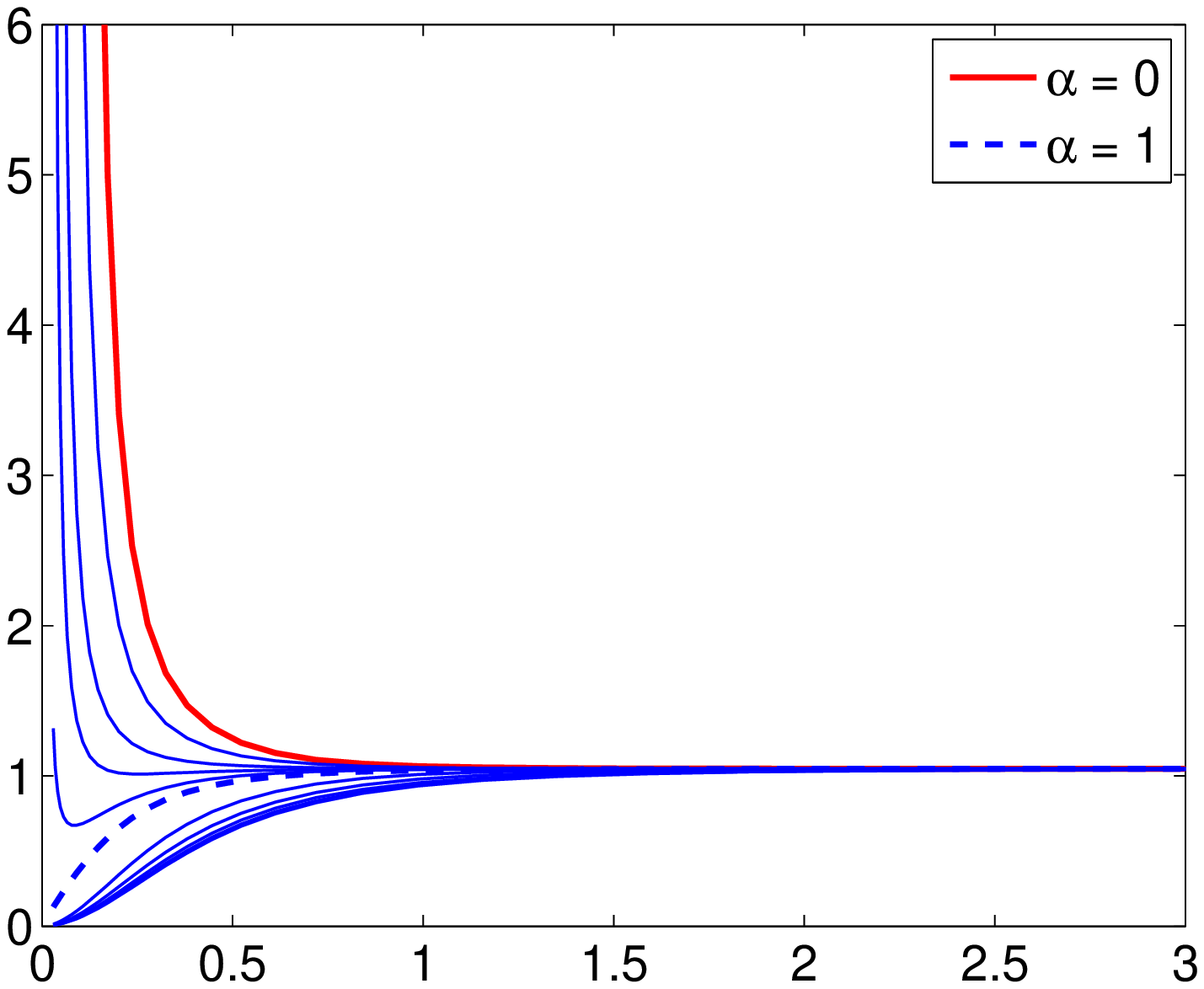,height=2 in}
\begin{picture}(0,0)
\put(-376,83){$\kappa$}
\put(-288,0){$\gamma$}
\put(-186,83){$\kappa$}
\put(-98,0){$\gamma$}
\end{picture}
}
\addtocounter{myfigure}{1}
\caption{{\it Relation of $\kappa$ and $\gamma$ for different values of 
$\alpha$ and $P_{\lambda}$}:
{\rm (a)} $P_{\lambda}=1$; {\rm (b)} $P_{\lambda}=0.75$; 
{\rm (c)} $P_{\lambda}=0.50$;
{\rm (d)} $P_{\lambda}=0.25$. }
\label{fig:gamma_kappa}
\end{figure}

As in the case of (\ref{equationalpha1dim1})--(\ref{equationalpha1dim2}),
we consider a frame of reference situated in the molecule of $B$,
i.e. molecules of $A$ diffuse with the diffusion constant 
$D_{A}+D_{B}$ and are removed in the ball around origin with probability
$P_\lambda$ during each time step. We use the dimensionless
parameters given by (\ref{defalpha}), (\ref{e:dim1}) and 
(\ref{defgamma}). Let $c_{k}(\hat{r})$ be the concentration 
of molecules of $A$ at the distance $\hat{r}$ from the origin. 
Each step of the algorithm changes the concentration which 
can be schematically described as follows:
$$
c_{k}(\hat{r})
\stackrel{[i]}{\longrightarrow}
c^{[i]}_{k}(\hat{r})
\stackrel{[ii]}{\longrightarrow}
c^{[ii]}_{k}(\hat{r})
\stackrel{[iii]}{\longrightarrow}
c_{k+1}(\hat{r}),
$$
where $c^{[i]}_{k}(\hat{r})$ (resp. $c^{[ii]}_{k}(\hat{r})$) 
is a concentration at the distance $\hat{r}$ 
from the origin after step [i] (resp. [ii]). 
Using the definition of steps [i]--[iii], we find
\begin{eqnarray}
c^{[i]}_{k}(\hat{r})
&=&
(1-P_{\lambda})\chi_{[0,1]}(\hat{r})c_{k}(\hat{r})+\chi_{(1,\infty)}(\hat{r})c_{k}(\hat{r}),
\label{step1itc}
\\
c^{[ii]}_{k}(\hat{r})
&=&
c_{k}^{[i]}(\hat{r}) + \omega \, \delta(\hat{r}-\alpha),
\label{step2itc}
\\
c_{k+1}(\hat{r})
&=&
\int^{\infty}_{0}
K(\hat{r},\hat{r}',\gamma)
c^{[ii]}_{k}(\hat{r}') \, \mbox{d} \hat{r}',
\label{step3itc}
\end{eqnarray}
where $\omega$ is a constant describing the production 
of molecules of $A$ in one time step and  
$K(z,z',\gamma)$ is a Green's function for the difusion 
equation given by 
$$
K(z,z',\gamma) 
=
\frac{z'}{z\gamma\sqrt{2\pi}}
\left(
\exp\left[-\frac{(z-z')^2}{2\gamma^{2}}\right]
-
\exp\left[-\frac{(z+z')^2}{2\gamma^{2}}\right] 
\right).
$$
Substituting (\ref{step1itc}) and (\ref{step2itc})
in (\ref{step3itc}), we obtain
$$
c_{k+1}(\hat{r}
)=(1-P_{\lambda})
\int^{1}_{0}
K(\hat{r},\hat{r}',\gamma) c_{k}(\hat{r}') 
\, \mbox{d}\hat{r}'
+
\int^{\infty}_{1}
K(\hat{r},\hat{r}',\gamma)
c_{k}(\hat{r}') 
\, \mbox{d}\hat{r}'
+\omega \, K(\hat{r},\alpha,\gamma).
$$
We are interested to find the fixed point $g(\hat{r})$ of this iterative 
scheme \cite{Erban:2009:SMR}. At steady state, the mass
lost in (\ref{step1itc}) is equal to the mass added
in (\ref{step2itc}), i.e. 
$4\pi \alpha^{2} \omega = P_{\lambda} \int^{1}_{0} g(z)4\pi z^{2} \, 
\mbox{d} z.$ Consequently, $g(\hat{r})$ satisfies the following equation
\begin{eqnarray}\label{A}
g(\hat{r})&=&(1-P_{\lambda})
\int^{1}_{0}
K(\hat{r},\hat{r}',\gamma) g(\hat{r}') 
\, \mbox{d}\hat{r}'
+
\int^{\infty}_{1}
K(\hat{r},\hat{r}',\gamma)
g(\hat{r}') 
\, \mbox{d}\hat{r}' \nonumber \\
&+& 
\frac{P_{\lambda} \, K(\hat{r},\alpha,\gamma)}{\alpha^{2}} 
\int^{1}_{0}
g(z) z^{2} \, \mbox{d} z.
\end{eqnarray}
Then the rate of removing of particles during one time step is
\begin{equation}\label{B}
\kappa=P_{\lambda}\int^{1}_{0}4\pi z^{2} g(z)dz.
\end{equation}
where $\kappa$ is the dimensionless reaction rate given by
\begin{equation}
\kappa=\frac{k_{1}\Delta t}{\newrho^{3}}.
\end{equation}
It is worth noting that $\kappa$ is defined with the help of
the time step $\Delta t$ and it is therefore different from
$\overline{\kappa}$ defined by (\ref{e:dim1}). In Figure \ref{fig:gamma_kappa},
we plot $\kappa$ as a function of $\gamma$, for different values 
of  probability $P_{\lambda}$ and ratio $\alpha$. 
Figure \ref{fig:gamma_kappa}(a) is calculated for $P_{\lambda}=1$, 
which corresponds to the Andrews and Bray model 
\cite{Andrews:2004:SSC}. Panels (b), (c) and (d) in Figure 
\ref{fig:gamma_kappa} correspond to $P_\lambda = 0.75$,
$P_\lambda = 0.5$ and $P_\lambda = 0.25$, respectively.
In each panel, the $\kappa$-$\gamma$ curves are plotted
for the values of ratio $\alpha$ equal to 
0, 0.5, 0.7, 0.8, 0.9, 1, 1.6, 2.5, 4, 6.3 and 10, 
starting always from the top in each panel. 
To solve equation \eqref{A} numerically, we use the condition 
$g(\hat{r}) \to 1$ as  $\hat{r} \to \infty$ 
to truncate the integrals to the finite domain 
\cite{Erban:2009:SMR}. The integrals over the finite domain are
then evaluated by the simpson rule.

\subsection{Probability of geminate recombination}
In Figure \ref{fig:gamma_kappa}, we observe that there
exist various combinations of the parameters $\gamma,$ $P_{\lambda}$ and 
$\alpha$ for which we obtain the same value of the dimensionless reaction 
rate $\kappa$. Even if we fix $\gamma$ (which is, roughly speaking,
equivalent to choosing the time step $\Delta t$), there are still
different choices of pairs $P_\lambda$ and $\alpha$ which lead
to the same reaction rate. 
For example, $\kappa=1$ and $\gamma=0.5$ can be achieved both for 
$P_{\lambda}=0.5 ,$ $\alpha= 4.4258$ and $P_{\lambda}=0.25 ,$ 
$\alpha= 0.8887$. As we observed in Section \ref{gemrecomb}, one 
possible way to distinguish different sets of parameters is by 
studying the geminate recombination probability. 
Let $p(\hat{r})$ be the probability that a molecule starting at 
$\hat{r}$ reacts before it escapes to infinity. It satisfies
the equation
\begin{equation}\label{e:gem_num}
p(\hat{r})
=
P_{\lambda} \int_{0}^{1} 
K(\hat{r},\hat{r}',\gamma) \, \mbox{d}\hat{r}'
+
(1-P_{\lambda})
\int_{0}^{1} 
K(\hat{r},\hat{r}',\gamma) \, p(\hat{r}') \, \mbox{d}\hat{r}'
+ 
\int_{1}^{\infty} 
K(\hat{r},\hat{r}',\gamma) \, p(\hat{r}') \, \mbox{d}\hat{r}',
\end{equation}
with the boundary condition 
\[
\lim_{\hat{r}\to \infty}p(\hat{r})=0. 
\]
The probability of geminate recombination is given
as $\phi = p(\alpha)$. Solving  \eqref{e:gem_num} numerically,
we find that the geminate recombination probability is 
$\phi=0.12$ for the first case 
($P_{\lambda}=0.5$, $\alpha= 4.4258$)
and  $\phi = 0.38$ for the second case 
($P_{\lambda}=0.25 $, $\alpha= 0.8887$)
which is a significant difference.

Another possibility to reduce the number of algorithm parameters
is by considering the values of realistic measurable parameters 
for a particular application. This will be shown in the following
section for the case of proteins.

\section{Illustrative Brownian dynamics results}
\label{secnumerresults}
In the previous sections, we derived relations between
the algorithm parameters $\newrho$, $\newsigma$, 
$\lambda$ (resp. $P_\lambda$) and the experimentally
measurable quantities. In this section, we illustrate
our results using a simple toy problem. We will consider
a cubic reactor of the size $L \times L \times L$ where
$L = 50$ nm. In the reactor, there are molecules of three 
chemical species $A$, $B$ and $C$ which are subject to
the reversible reaction (\ref{reversiblereaction}). The
molecules diffuse inside the reactor. The boundary of
the reactor is considered to be non-reactive (reflective)
and we start with 5 molecules of each species in the
domain.

\begin{figure}
\centerline{
\hskip 6mm
\raise 2in \hbox{\raise 0.9mm \hbox{(a)}}
\hskip -8mm
\epsfig{file=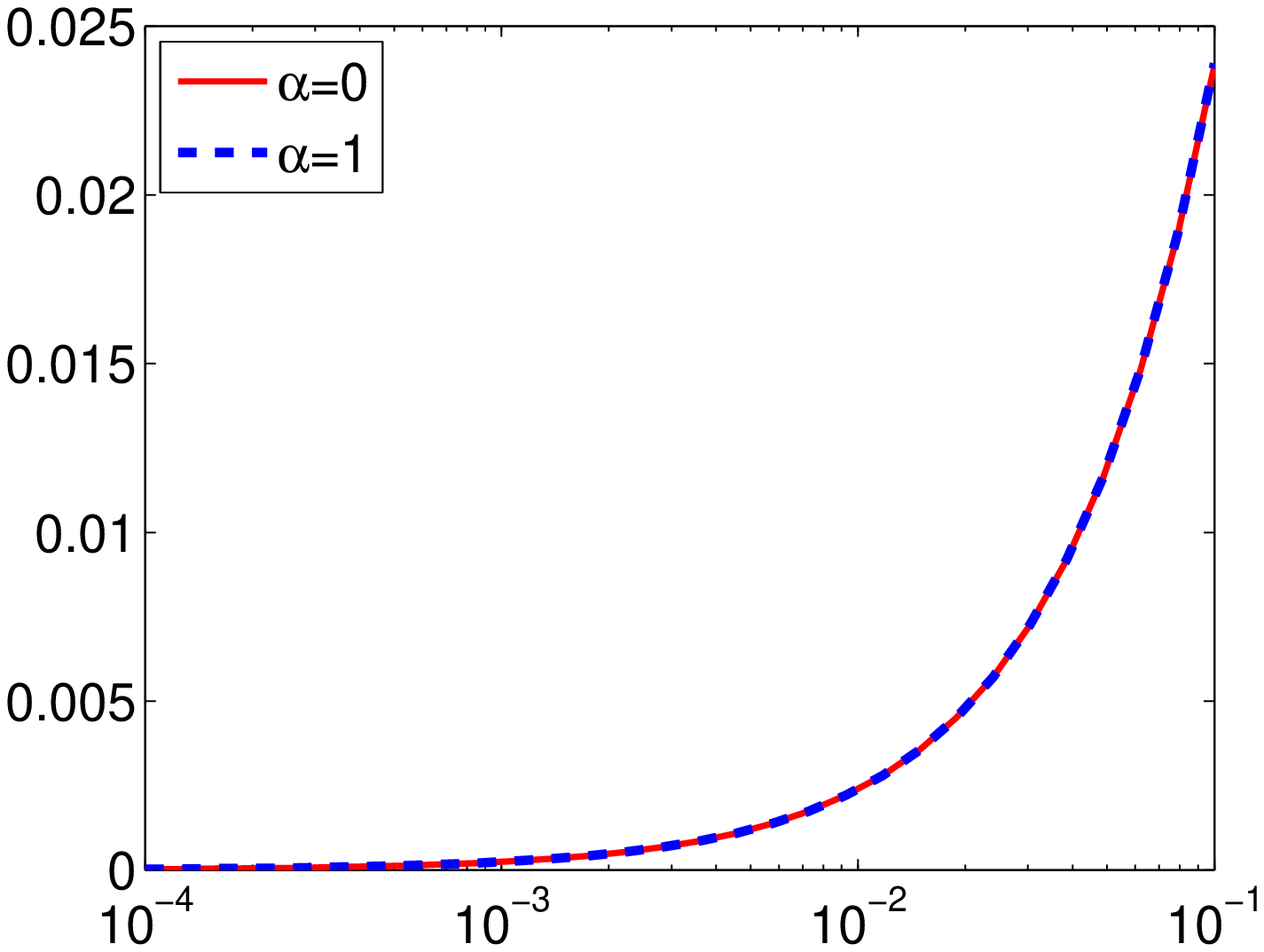,height=2in}
\raise 2in \hbox{\raise 0.9mm \hbox{(b)}}
\hskip -8mm
\epsfig{file=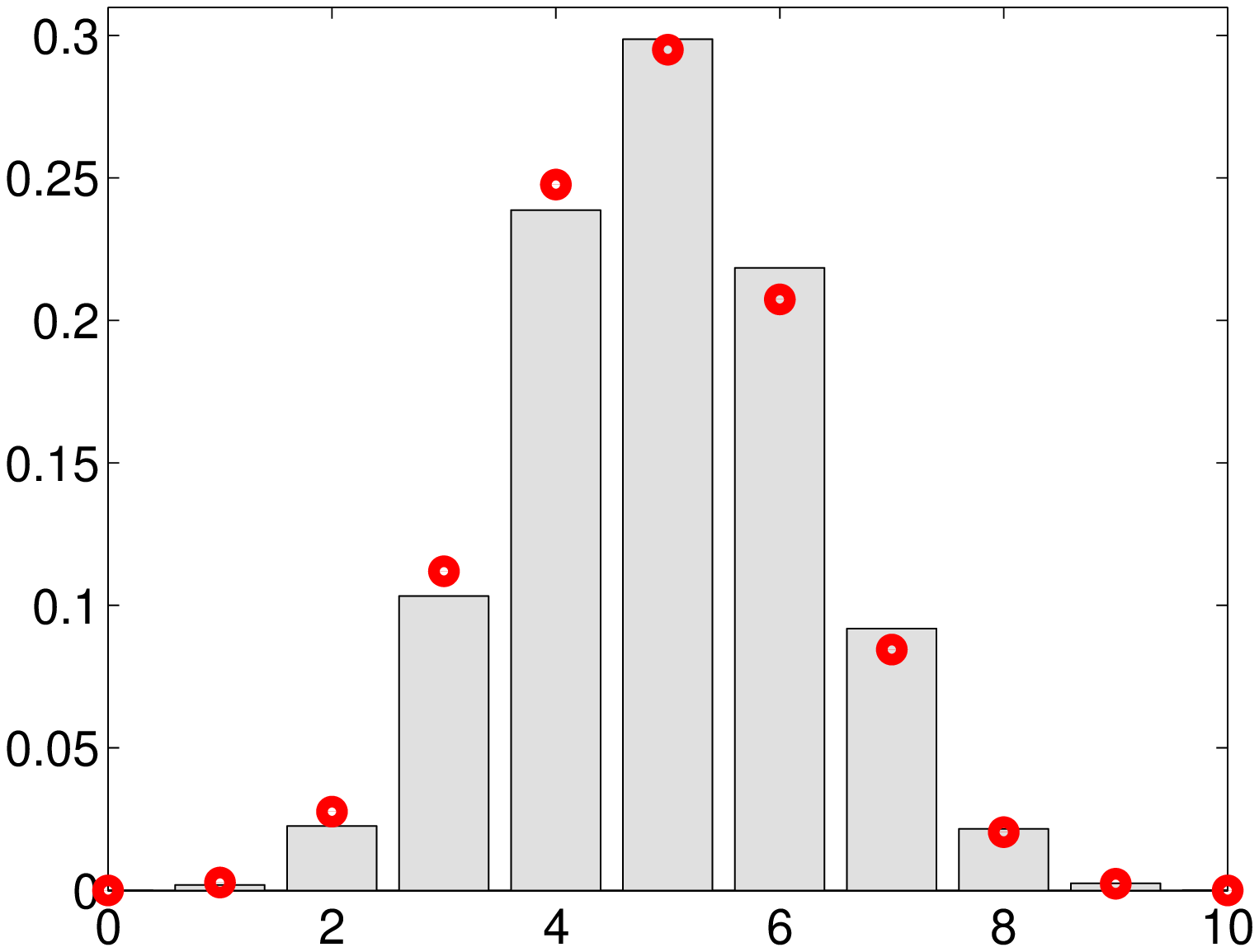,height=2in}
\begin{picture}(0,0)
\put(-386,97){$P_{\lambda}$}
\put(-288,0){$\kappa$}
\put(-196,27){\begin{sideways}stationary distribution\end{sideways}}
\put(-150,-2){number of molecules of $A$}
\end{picture}
}
\addtocounter{myfigure}{1}
\caption{{\rm (a)} {\it  $P_{\lambda}$ as a function of  
$\kappa$ for  $\alpha=0$ and $\alpha=1$, and $\gamma=1$.} 
{\rm (b)}
{\it Stationary distibution of molecules of $A$ computed by
the Brownian dynamics simulation for  
$P_{\lambda}=4.95 \times 10^{-5}$ and $\alpha=1$.}}
\label{fig:P_sigma}
\end{figure}

Using typical diffusion constants of proteins
$D_{A}=D_{B}=D_{C}=10^{-5}$ cm$^{2}$ s$^{-1}$, 
the reaction radius $\bar{\rho}=2$ nm and 
the time step $\Delta t=10^{-9}$ s, we obtain
that the dimensionless parameter $\gamma$ defined
by (\ref{defgamma}) is $\gamma = 1$. Considering
that typical rate constants of protein-protein interactions
are about $10^{6}$ M$^{-1}$ s$^{-1}$, we obtain
that the dimensionless parameter $\kappa$ is of the
order $10^{-4}$. In Figure~\ref{fig:P_sigma}(a), 
we plot the dependence of the probability $P_{\lambda}$ as 
a function of $\kappa$ for $\alpha=0$ and $\alpha=1$ in 
the case where $\gamma=1$. 
Note that in the case $\alpha=0$, the equation \eqref{A} becomes
\begin{eqnarray*}
g(\hat{r})&=&(1-P_{\lambda})
\int^{1}_{0}
K(\hat{r},\hat{r}',\gamma) g(\hat{r}') 
\, \mbox{d}\hat{r}'
+
\int^{\infty}_{1}
K(\hat{r},\hat{r}',\gamma)
g(\hat{r}') 
\, \mbox{d}\hat{r}' \nonumber \\
&+& 
\frac{P_{\lambda}}{4\pi\gamma^{3}} 
\sqrt{\frac{2}{\pi}} 
    \exp \left( -\frac{\hat{r}^{2}}{2\gamma^{2}} \right)  
\int^{1}_{0}
g(z) z^{2} \, \mbox{d} z.
\end{eqnarray*}
As we can see, the probability $P_{\lambda}$ appears to 
be independent of $\alpha$ for this particular parameter range 
of $\kappa$. We thus set $\alpha=1$, i.e. $\newsigma=\newrho$.
We use $k_{1}=10^{6}$ M$^{-1}$ s$^{-1}$ and $k_2 = 66.7$
s$^{-1}$.
Then equations (\ref{A})--(\ref{B}) imply that 
$P_{\lambda}=4.95 \times 10^{-5}$
and we can use the steps [i]--[iii] to simulate the illustrative
toy model. If the diffusive step [iii] places a molecule outside
the reactor, we return it back using mirror reflection.
This is a typical way to implement no-flux boundary conditions.
For discussion of more complicated boundary conditions, see
\cite{Erban:2007:RBC}.

To visualize the results of stochastic simulation, we compute
the stationary distribution of the numbers of molecules of $A$ 
in the whole reactor as follows. We run the simulation for a 
long time and we record the number of molecules of $A$ at equal 
time intervals. The resulting (grey) histogram is plotted in  
Figure \ref{fig:P_sigma}(b). Since the domain is relatively
small, we can make a direct comparison with the stationary
histogram obtained by the (spatially-homogeneous, well-mixed) 
simulation of the reversible reaction (\ref{reversiblereaction})
by the Gillespie SSA \cite{Gillespie:1977:ESS}, which is equivalent
to solving the corresponding chemical master equations. The results
are plotted as red circles in Figure \ref{fig:P_sigma}(b). 
As expected, the comparison with the Brownian dynamics
(spatial stochastic simulation) is excellent.

\subsection{Geminate recombination}
In our second illustrative example, we use the stochastic
simulation of $\lambda$-$\newrho$ model to directly validate
our formulae for geminate recombination. We simulate
the behaviour of molecules of $A$, $B$ and $C$ in the cubic
reactor as before. Whenever two molecules of $A$ and $B$ are 
introduced in the system, we check if they react with each 
other again before reacting with another molecule or hitting 
the boundary of the reactor. We then approximate the geminate 
recombination probability, by the  ratio of geminate recombination 
events over the total number of forward reactions 
(\ref{irreversiblereaction}) occurring in the simulation. 

\begin{figure}
\centerline{
\hskip 6mm
\raise 2in \hbox{\raise 0.9mm \hbox{(a)}}
\hskip -8mm
\epsfig{file=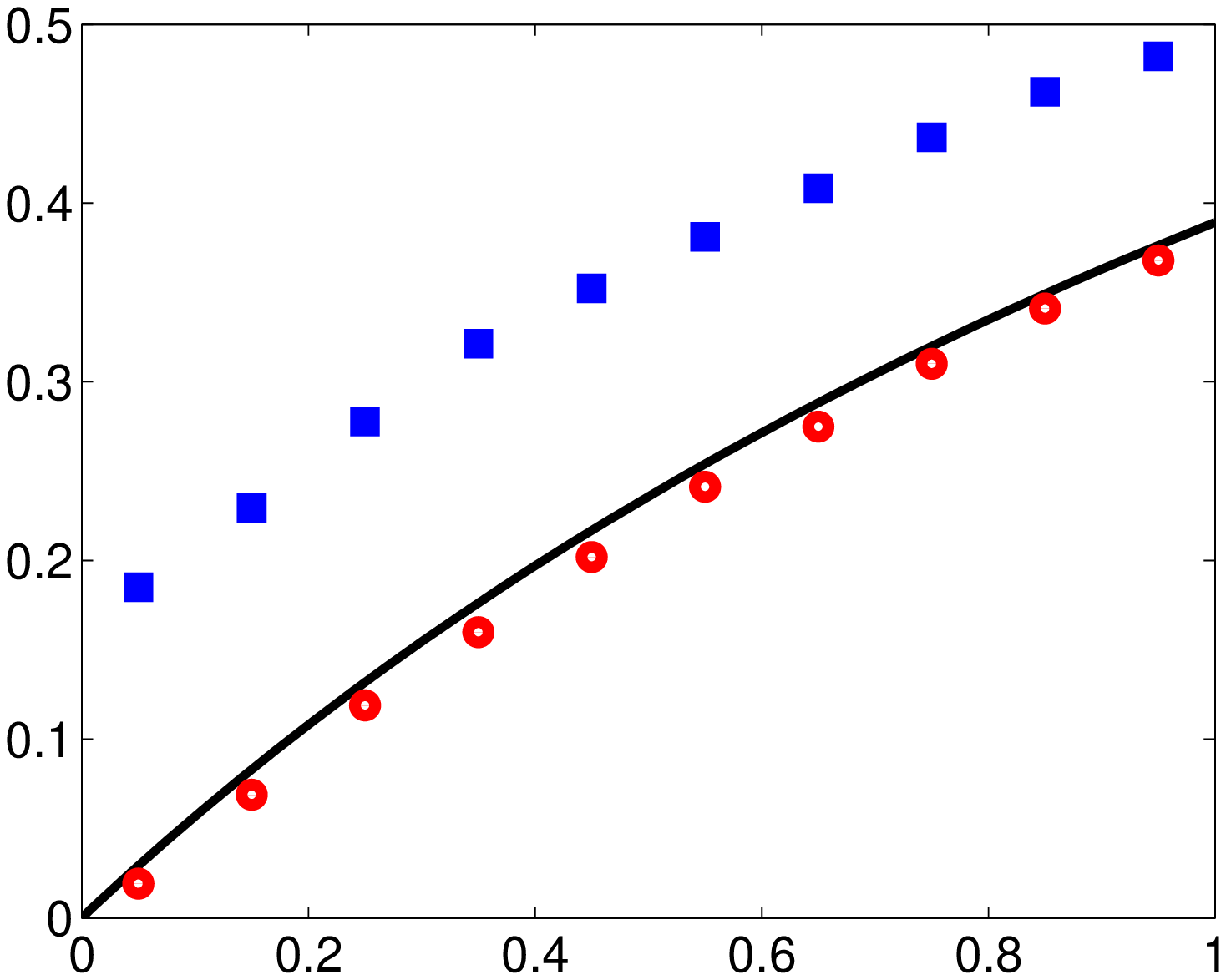,height=2in}
\raise 2in \hbox{\raise 0.9mm \hbox{(b)}}
\hskip -8mm
\epsfig{file=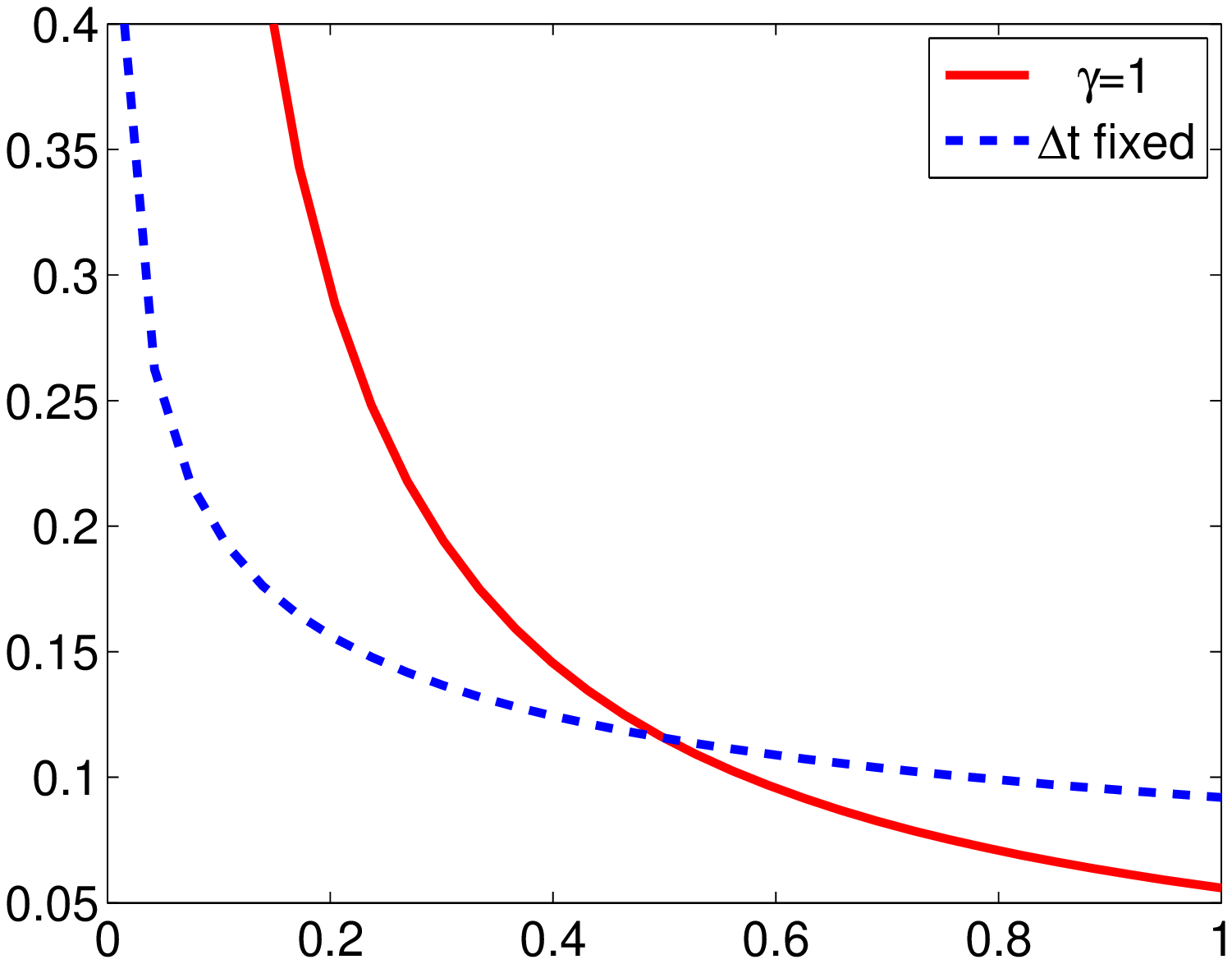,height=2in}
\begin{picture}(0,0)
\put(-382,95){$\phi$}
\put(-288,0){$P_{\lambda}$}
\put(-195,95){$\newrho$}
\put(-98,0){$P_{\lambda}$}
\put(-143,120){$\gamma=1$}
\put(-85,85){$\gamma >1$}
\put(-160,26){$\gamma <1$}
\end{picture}
}
\addtocounter{myfigure}{1}
\caption{(a) 
{\it Comparison of \eqref{e:gem_num} with the spatial 
stochastic simulations algorithm for $\gamma=1$.}
(b) {\it  Dependence of $\newrho$ on $P_{\lambda}$ 
calculated for $\gamma=1$ (red solid line)
and for $\Delta t= 33 \times 10^{-4}$ s (blue dashed line).}
}
\label{fig:change_rhoP_sigma1}
\end{figure}

Solving \eqref{e:gem_num} for the parameters used in Figure 
\ref{fig:P_sigma}(b), we find that $\phi=2.45 \times 10^{-5}$
which is negligible. In order to illustrate the strength
of the formula \eqref{e:gem_num}, we will use different parameter 
values for which the gemination combination probability is
significant, namely
$D_{A}=D_{B}=D_{C}= 1\;\mu\mbox{m}^{2}\,\mbox{sec}^{-1},$ rate constants 
$k_{1}=1\;\mu\mbox{m}^{3}\,\mbox{sec}^{-1},$ 
$k_{2}=0.005\;\mbox{sec}^{-1}$, $L=20 \; \mu \mbox{m}$, $\alpha=0$,
$\gamma=1$ and different values for the probability $P_{\lambda}$. 
In Figure \ref{fig:change_rhoP_sigma1}(a), we compare
the results obtained by \eqref{e:gem_num} with the results
estimated from the Brownian dynamics simulations (red circles). The comparison
is very good. We also plot the results estimated from the same
stochastic simulation showing how often the molecules of $A$ and 
$B$ which were created from the same complex $C$ react with each
other (blue squares). The difference between the (red) circles
and (blue) squares is that in the former we do not consider the
event to be a geminate recombination if either of the  molecules of $A$ or
$B$ has hit the domain boundary, before they react again with each other. 
Thus (blue) squares give an upper estimate of the
geminate recombination given by \eqref{e:gem_num}, because we
have finite number of molecules in the box (on average only 5
molecules). In particular the blue squares would approach the theory and the red circles for simulations of a large number of molecules. 

In Figure \ref{fig:change_rhoP_sigma1}(a), we fixed the value of
$\gamma$ as 1, since this is the value for which the spatial 
stochastic simulation algorithm discussed in Section \ref{secnumer}
is the most relevant. In particular, every time we change the probability 
$P_{\lambda}$, we also change the time step $\Delta t$ and the 
reaction radius $\newrho$. In Figure \ref{fig:change_rhoP_sigma1}(b), 
we present the dependence of the binding radius $\newrho$ on 
$P_{\lambda}$. Each point on this curve corresponds to a different
time step. Another option to compare the results would be to
choose $\Delta t$ to be fixed for all the different probabilities 
$P_{\lambda}$, which means that $\gamma$ would have to be different 
in every simulation. The dependence of the binding radius
$\newrho$ on $P_{\lambda}$ for fixed $\Delta t$ is also plotted in 
Figure \ref{fig:change_rhoP_sigma1}(b) for comparison.
We choose $\Delta t= 33 \times 10^{-4}$ s, which 
is the value for which $\gamma=1$, when $P_{\lambda}=0.5$. 
As we can see in both cases the binding 
radius $\newrho$ is a decreasing function of the probability $P_{\lambda}$. When we keep $\Delta t$ fixed, 
$\newrho$ decreases slower than it does in the case of fixed 
$\gamma$, which implies that $\gamma$ in this case of fixed 
$\Delta t$ becomes smaller than $1$ as $P_{\lambda}$ gets 
smaller than $0.5$. 

\section{Discussion}

Several algorithms for stochastic simulation of reaction-diffusion 
processes in cell and molecular biology have been proposed in the 
literature. Some of these methods are lattice-based and can be 
equivalently described in terms of the reaction-diffusion master 
equation (RDME) \cite{Hattne:2005:SRD,Isaacson:2006:IDC}.
Approaches to simulate RDME-based models efficiently have been
recently proposed \cite{Drawert:2010:DFS,Ferm:2010:AAS} and the 
RDME methods were generalized to unstructured meshes 
\cite{Engblom:2009:SSR}, but other open questions remain. 
For example, the relation of RDME to more detailed off-lattice 
models \cite{Isaacson:2009:RME,Erban:2009:SMR} and
efficient ways to investigate the dependence of 
simulation results on the model parameters, e.g. 
efficient bifurcation analysis of stochastic models 
\cite{Qiao:2006:SDS}. 

In this paper, we studied an alternative approach to stochastic
reaction-diffusion modelling. We presented a class of Brownian 
dynamics algorithms. These algorithms are 
off-lattice and can, in principle, provide more details.
However, they share some problems with the RDME-based simulations,
e.g. all stochastic models are usually more computationally intensive
than solving the corresponding deterministic reaction-diffusion
partial differential equations. 
One way to decrease the computational intensity
is to consider Brownian dynamics of point-like particles
\cite{Andrews:2004:SSC}. In \cite{Erban:2009:SMR}, we presented
$\lambda$-$\newrho$ approach which provides more flexibility in choosing 
the reaction radius $\newrho$ than one-parameter based models.
In this paper, we show that this approach can be generalized
to the case of reversible reactions, addressing the criticism
mentioned in the recent paper describing the Smoldyn algorithm 
\cite{Andrews:2010:DSC} (page 5). In particular, we show that, 
in the parameter regime relevant to protein-protein simulation, 
it is possible to avoid the concept of the unbinding radius 
$\newsigma$. We illustrate that the same results can be obtained
for $\newsigma=0$ and for $\newsigma=\newrho$. If we consider smaller 
reaction radii or larger reaction rates, then the unbinding radius 
has to be taken into account. We derived formulae for the probability
$\phi$ of geminate recombination
which can be used to select the appropriate algorithm parameters.
In particular, we also generalized the results of Andrews and
Bray \cite{Andrews:2004:SSC} (which were derived for $P_\lambda=1$),
to the case of arbitrary reaction probability $P_\lambda \in (0,1]$.
It is worth noting that the RDME-based approaches do not have special 
difficulties
with simulating reversible reactions, because they can be implemented
as two reactions (\ref{irreversiblereaction}) and (\ref{reversestep})
in a straightforward way. 

Bimolecular reactions are very common in cell biology 
\cite{Marianayagam:2004:PTP,Alberts:2002:MBC} and therefore, 
it is important to study their correct implementation 
in the computational algorithms \cite{Erban:2009:SMR}. 
However, there are several other issues which needs to be 
considered in order to simulate realistic spatially-distributed 
reaction-diffusion systems \cite{Tostevin:2007:FLP}.
Brownian dynamics require extra attention when simulating 
reactive boundaries (e.g. reactions on the plasma membrane) 
\cite{Erban:2007:RBC,Andrews:2009:APS} and one should 
also have in mind steric interactions, i.e. the consequences of 
macromolecular crowding inside the cytoplasm \cite{Alberts:2002:MBC}. 
We will address this issue in a future publication.

\vskip 2mm
\noindent
{\bf Acknowledgements.} This publication is based on work (JL, KZ, RE) 
supported by Award No. KUK-C1-013-04, made by King Abdullah 
University of Science and Technology (KAUST). RE was also supported
by the European Research Council Starting Independent Researcher
Grant and by Somerville College, Oxford.

\bibliographystyle{siam}

\end{document}